\documentclass[useAMS,usenatbib]{mn2e}
\usepackage{amsmath}
\usepackage{amssymb}

\usepackage{graphics}
\usepackage{graphicx}
\usepackage{epstopdf}
\usepackage{footnote}
\usepackage{tablefootnote}
\usepackage{txfonts}
\usepackage{lipsum}
\usepackage{float}
\usepackage[draft]{hyperref}
\usepackage{macros}

\makesavenoteenv{tabular}

\DeclareSymbolFont{cmletters}{OML}{cmm}{m}{it}
\DeclareMathSymbol{v}{\mathalpha}{cmletters}{"76}

\usepackage[usenames,dvipsnames,svgnames,table]{xcolor}
\definecolor{darkblue}{rgb}{0.0,0.0,0.3}
\hypersetup{colorlinks,breaklinks,
            linkcolor=darkblue,urlcolor=darkblue,
            anchorcolor=darkblue,citecolor=darkblue}

\title[Modeling Sagittarius A*]{The Disc-Jet Symbiosis Emerges: Modeling the Emission of Sagittarius A* with Electron Thermodynamics
 }
\author[S. M. Ressler, A. Tchekhovskoy, E. Quataert, C. F. Gammie]{S. M. Ressler$^{1},$ A. Tchekhovskoy$^{1}$\thanks{Einstein and TAC Fellow}, E. Quataert$^{1}$, C. F. Gammie$^{2,3}$\\
$^{1}$Departments of Astronomy \& Physics, Theoretical Astrophysics Center, University of California, Berkeley, CA 94720  \\
$^{2}$Department of Astronomy, University of Illinois, 1002 West Green Street, Urbana, IL 61801\\
$^{3}$Department of Physics, University of Illinois, 1110 West Green Street, Urbana, IL 61801}

\begin{document}
\pagerange{\pageref{firstpage}--\pageref{lastpage}} \pubyear{2016}
\maketitle
\label{firstpage}
\begin{abstract}
  We calculate the radiative properties of Sagittarius A* -- spectral energy distribution, variability, and radio-infrared images -- using the first 3D, physically motivated black hole accretion models that directly evolve the electron thermodynamics in general relativistic MHD simulations.    These models reproduce the coupled disc-jet structure for the emission favored by previous phenomenological analytic and numerical works.   More specifically, we find that the low frequency radio emission is dominated by emission from a polar outflow while the emission above $100$~GHz is dominated by the inner region of the accretion disc. 
The latter produces time variable near infrared (NIR) and X-ray emission, with frequent flaring events (including IR flares without corresponding X-ray flares and IR flares with weak X-ray flares).  
The photon ring is clearly visible at $230$~GHz and $2$~microns, which is encouraging for future horizon-scale observations.
We also show that anisotropic electron thermal conduction along magnetic field lines has a negligible effect on the radiative properties of our model.
We conclude by noting limitations of our current generation of first-principles models, particularly that the outflow is closer to adiabatic than isothermal and thus underpredicts the low frequency radio emission.
 \end{abstract}

\begin{keywords}
MHD --- galaxy: centre --- relativistic processes --- accretion --- black hole physics
\end{keywords}
\section{Introduction}
Sagittarius A* (Sgr A*), the supermassive black hole at the center of our galaxy, is a prime candidate for directly comparing general relativistic magnetohydrodynamic (GRMHD) simulations of accretion discs to observations.   Not only is there a wealth of observational data in the radio-millimetre \citep{Falcke1998,An2005,Doeleman2008,Bower2015}, near-infrared \citep{Genzel2003,Do2009,Schodel2011}, and X-ray \citep{Baganoff2003,Neilsen2013} bands, but the Event Horizon Telescope \citep{Doeleman2008} and GRAVITY \citep{Gillessen2010} will soon be able to spatially resolve the structure of the innermost region of the disc near the event horizon.  

The accretion rate in Sgr A* is orders of magnitude less than the Eddington limit, putting it in the Radiatively Inefficient Accretion Flow (RIAF) regime, characterised by a geometrically thick, optically thin disc \citep{Ichimaru1977,Rees1982,1994ApJ...428L..13N,2001ASPC..224...71Q,2014ARA&A..52..529Y}.  This particular class of accretion discs in some ways lends itself well to numerical simulation, given the dynamical unimportance of radiation and the large scale height of the disc that can be more easily resolved.  Over the past few decades, several numerical methods to simulate single-fluid RIAFs around rotating black holes in full general relativity have been developed (e.g. \citealt{Komiss1999,DeVilliers2003,Gammie2003,Sasha2007,White2016}).

On the other hand, the low densities typical of RIAFs imply that the electron-ion Coulomb collision time is much longer than an accretion time, so a single fluid model of the thermodynamics is not applicable. However, in the limit that the electrons are colder than the protons, $T_e \lesssim T_p$, which is generally expected for RIAFs, these single-fluid simulations should provide a reasonable description for the total gas properties.  Thus, to first approximation, the accretion dynamics, magnetic field evolution, and ion thermodynamics are known but the electron temperature is undetermined.  Previous approaches to modelling the emission from single-fluid RIAF simulations have attempted to overcome this limitation by adopting simplified prescriptions for the electron thermodynamics, such as taking $T_e/T_p = {\rm const.}$ (e.g. \citealt{Mosci2009}), splitting the simulation into jet and disc regions with different electron temperatures in each (e.g. \citealt{Mosci2014,CK2015}), or by solving a 1D, time-independent electron entropy equation in the midplane and interpolating to the rest of the grid (e.g. \citealt{Shcherbakov2012}).  Recently, however, we have developed a model which allows for the self-consistent evolution of the electron entropy alongside the rest of the GRMHD evolution, including the effects of electron heating and electron thermal conduction along magnetic field lines \citep{Ressler2015}. This model has been further extended by \citet{Sadowski2016} to include the dynamical effects of radiation and Coulomb Collisions on the fluid (while neglecting electron conduction), where they demonstrate that these effects are negligible for the accretion rate of Sgr A*; thus we neglect them here.

Here we present the observational application of that electron model to Sgr A* using 3D GRMHD simulations.  Throughout we focus on emission by thermal electrons. The aim of this work is to elucidate the basic properties of a fiducial model that is representative of simulations that include our electron entropy evolution.  We do not provide an exhaustive study of parameter space in order to find a ``best-fit'' model.  This is in part because we believe that the theoretical problem in its present state is too degenerate and uncertain to warrant such inferences. We do, however, compare and contrast our results to observations of Sgr A* and previous models.

The literature has used various terms to distinguish between types of outflow in black hole accretion disc systems. Most notable are the labels``jet,'' ``disc-jet,'' and ''wind,''  (see section 3.3 in \citealt{2014ARA&A..52..529Y} for a review).  ``Jet'' typically refers to the \citet{BZJet} model, which describes an electromagnetically dominated, relativisitic outflow powered by the spin of the black hole.  In GRMHD simulations, the thermodynamics are unreliable in this region due to its high magnetization.  Thus we do not attempt to model the emission from the jet but exclude it from the domain when calculating the spectra (see \S \ref{sec:rad} for details).  The ``disc-jet'' is the label typically given to the more mildly relativistic outflow sourced by the accretion disc (e.g., the \citealt{BPJet} or \citealt{LyndenBell2003} models, see \citealt{Yuan2015} for the distinction), while the term ``wind'' generally refers to non-relativistic outflow that occupies a larger solid angle. The thermodynamics of these regions are more reliably captured by GRMHD simulations since they are not as extremely magnetized.  In the present work we do not make a precise distinction between the labels``disc-jet'' and the ``wind,'' but will generally use the term ``outflow'' and ``disc-jet'' to refer to the disc-jet and wind regions.

The paper is organized as follows.  \S \ref{sec:fluid} describes our GRMHD method for electron entropy evolution, \S \ref{sec:rad} describes how we construct spectra and images of Sgr A*, \S \ref{sec:fid} describes the basic parameters and initial conditions of our fiducial model, \S \ref{sec:res} presents the results, \S \ref{sec:Thermpol} discusses the thermodynamics of the outflowing polar regions, \S \ref{sec:comp} compares our model to the phenomenological disc-jet models in the literature, and \S \ref{sec:conc} concludes.

For convenience, we absorb a factor of $\sqrt{4 \pi}$ into the definition of the magnetic field 4-vector, $b^\mu$, so that the magnetic pressure is $P_m = b^2/2$.  Furthermore, we set $GM=c  = 1$ throughout, where $G$ is the gravitational constant, $M$ is the black hole mass, and $c$ is the speed of light.

\begin{figure*}
    \includegraphics[scale = .08]{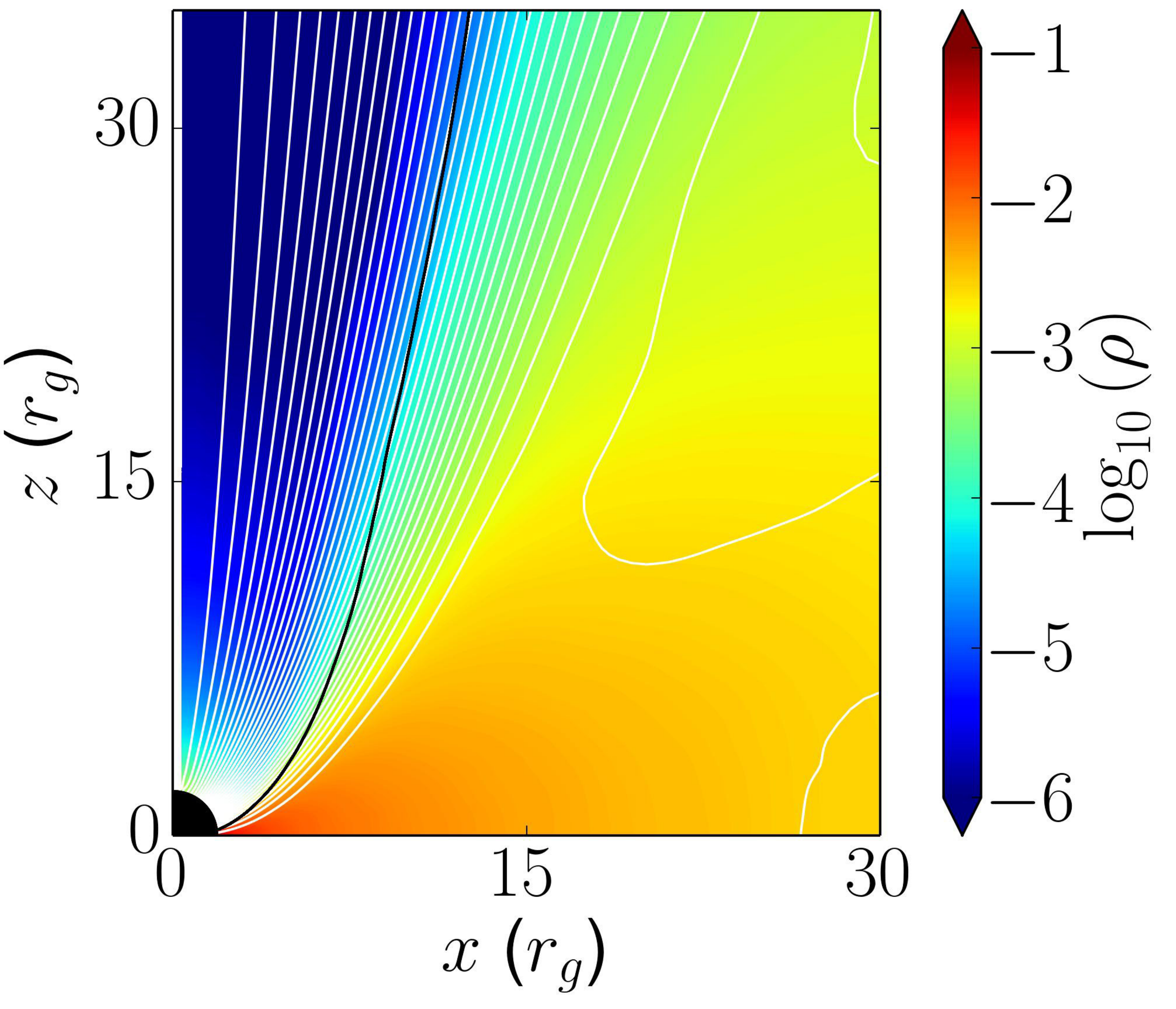} 
     \includegraphics[scale = .08]{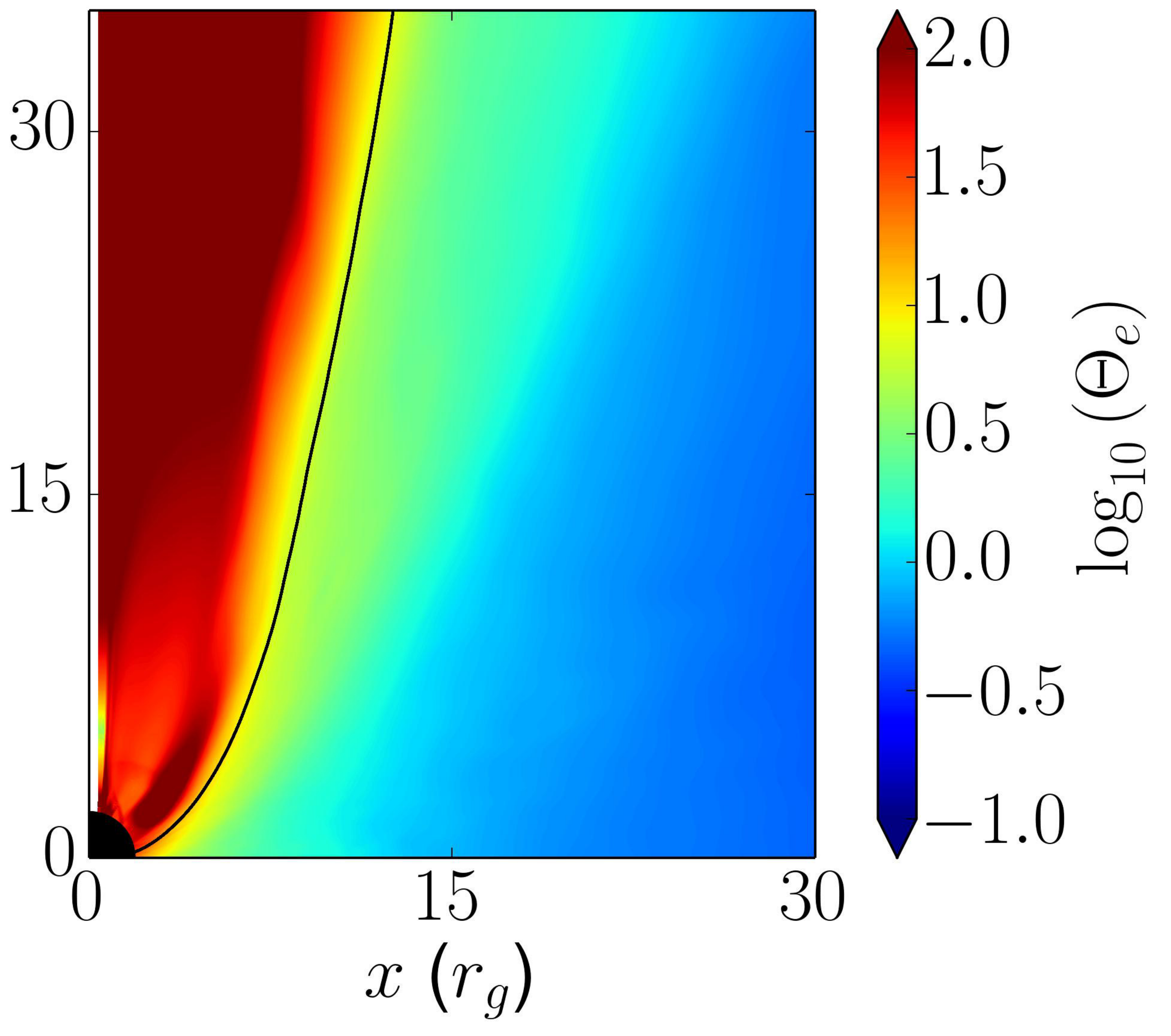}
     \includegraphics[scale = .08]{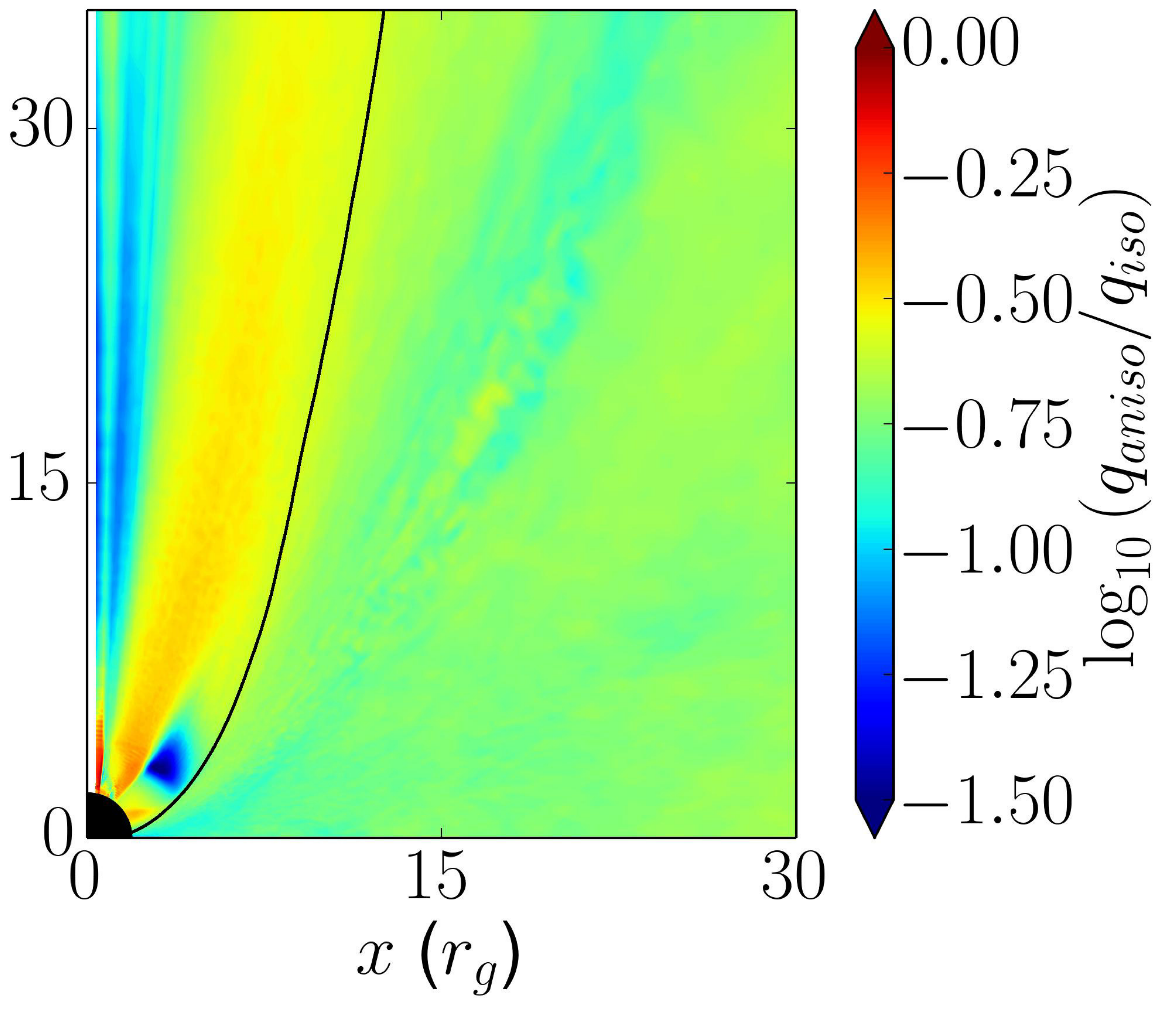}
         \includegraphics[scale = .08]{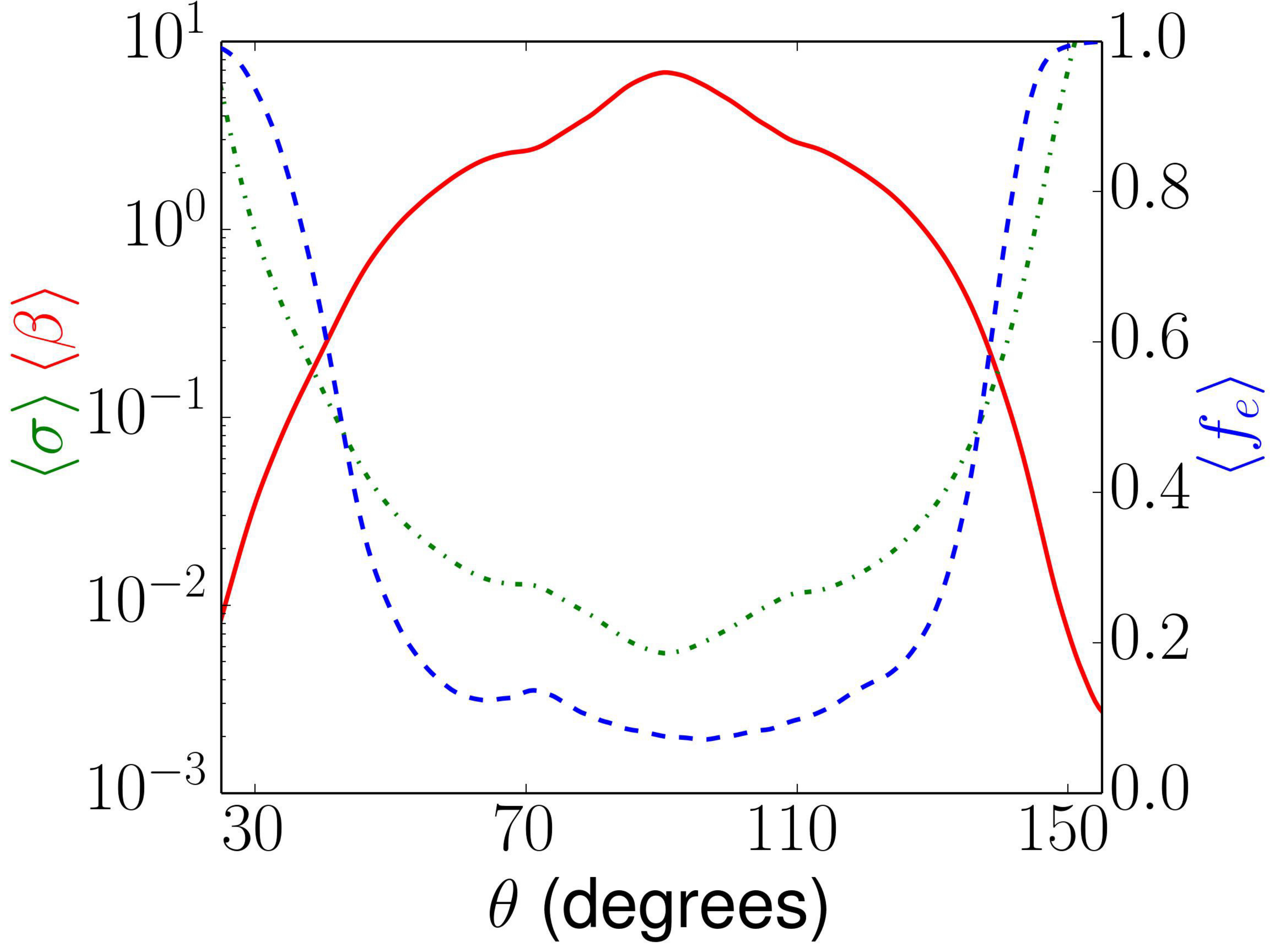} {\color{white} space}
                  \includegraphics[scale = .08]{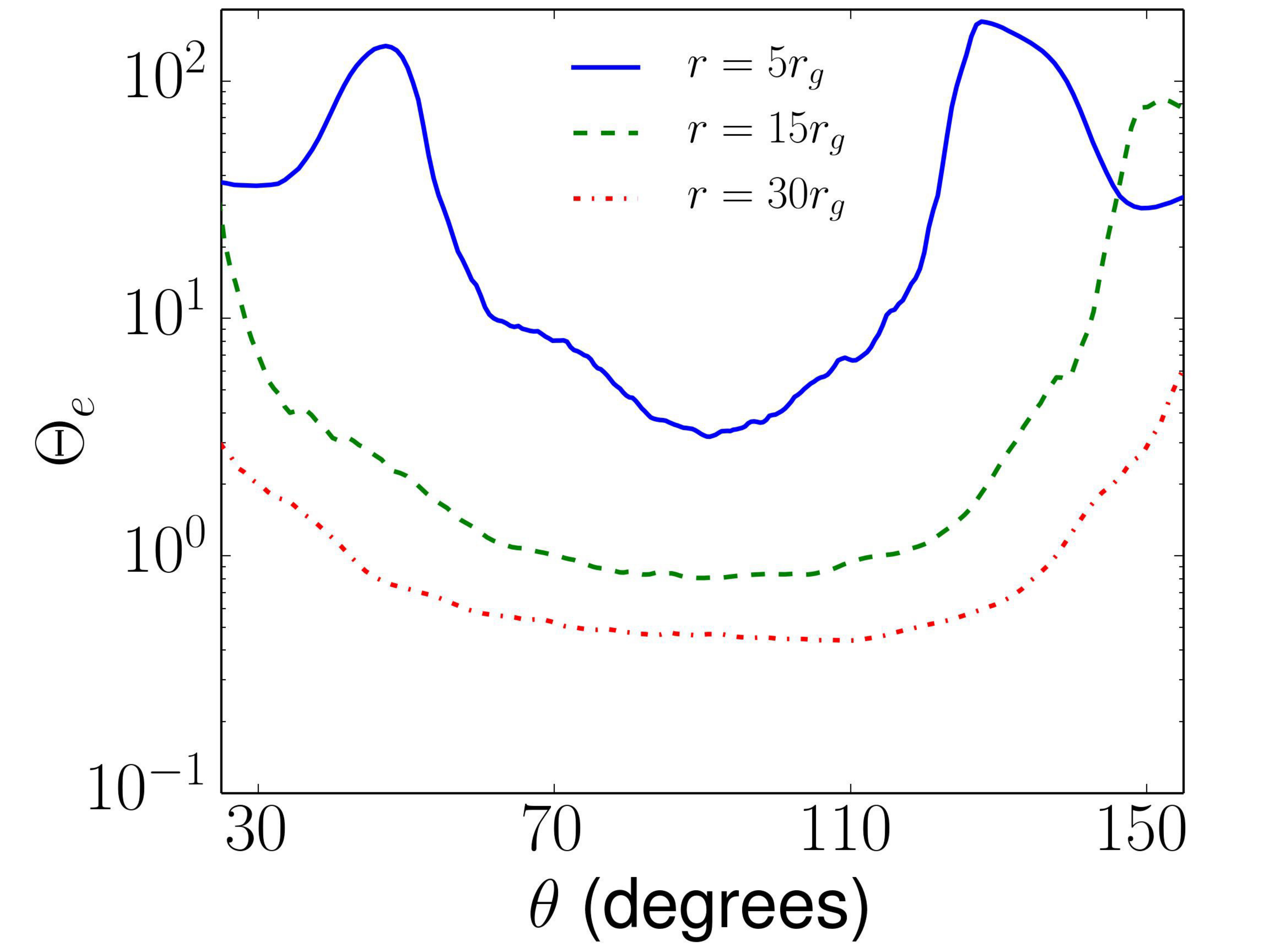}

\caption{Properties of our 3D black hole accretion simulations. The top left panel shows the density over-plotted with white magnetic field lines, the top middle panel shows the time- and $\varphi$-average electron temperature in units of $m_e c^2$, and the top right panel shows the ratio of the anisotropic (field-aligned) heat flux to the isotropic heat flux (both computed from a simulation without electron conduction).  All quantities in the top panel have been folded across the equator, and black lines denote the $b^2/\rho =1$ contour.  The bottom left panel shows the angular variation of the plasma parameter, $\beta \equiv 2P_{g}/b^2$, the magnetization parameter, $\sigma \equiv b^2/\rho$, and the electron heating fraction, $f_e$, averaged over $r$ from the event horizon to $25 r_g$, while the bottom right panel shows the angular variation of the electron temperature at 5, 15, and 30 $r_g$.  All quantities have been averaged over $\varphi$ and time from $15,000-19,000$ $r_g/c$.  Note how the relativistic electron temperatures important for synchrotron emission are strongly concentrated in the coronal and outflowing regions where $\beta \lesssim 1$.  This is a consequence of the strong $\beta$-dependence of our electron heating fraction, $f_e$ (see \S \ref{sec:fluid}). }%
\label{fig:SANEFluid}
\end{figure*}

\section{Fluid Model and Electron Thermodynamics}
\label{sec:fluid}
Using the assumption that $T_e \lesssim T_p$ we take the solution of the single-fluid, ideal GRMHD equations to be a good approximation for the total fluid number density, $n$, pressure, $P_g$, magnetic field four-vector, $b^\mu$, and four-velocity, $u^\mu$.  We solve these equations using a version of the numerical code {\tt HARM} \citep{Gammie2003} that we parallelized using message passing interface (MPI), extended to 3D, and made freely available online as the {\tt HARMPI} code.\footnote{\url{https://github.com/atchekho/harmpi}}  For the electron variables, we use the charge neutrality  assumption to constrain the electron number density and four-velocity to be the same as that of the ions (i.e.; $n_e = n_i = n, u^\mu_e = u^\mu_i = u^\mu$) but evolve a separate entropy equation to solve for the electron temperature:
\begin{equation}
  \rho T_e u^\mu \partial_\mu s_e = f_e  Q - \nabla_\mu q_e^\mu - a_\mu q_e^\mu
\end{equation}
where $f_e$ is a function of the local plasma parameters determining the fraction of the total heating rate per unit volume ($Q$) given to the electrons,  $q_e^\mu = \phi \hat b^\mu$ is the anisotropic thermal heat flux along field lines, $\hat b^\mu$ is a unit vector along $b^\mu$, and $a^\mu = u^\nu \nabla_\nu u^\mu$ is the four acceleration. The latter properly accounts for gravitional redshift of the heat flux \citep{ManiModel}. To calculate $Q$, we directly compare the internal energy obtained from solving an entropy conserving equation to the total internal energy of the gas as described in \citet{Ressler2015}. As in all conservative GRMHD codes, the heating is provided by grid-scale dissipation that is a proxy for of magnetic reconnection, shock heating, Ohmic heating, and turbulent damping. In this work, we determine $f_e$ via equations (48) and (49) in \citet{Ressler2015}, which were obtained from a fit to plasma heating calculations \citep{Howes2010} and are reasonably accurate at modelling particle heating in the solar wind \citep{Howes2011}.  The key qualitative feature of this prescription for $f_e$ is that it depends on the plasma $\beta$-parameter, $\beta \equiv P_g/P_m$, the ratio between the fluid and magnetic pressures: electrons (ions) are predominantly heated for $\beta \lesssim 1$ $(\beta \gtrsim 1)$, which is a general result predicted by linearizing the Vlasov equation and calculating the fractional heating rates of the two species due to MHD turbulence \citep{Quataert1999}. 

Thus, for a magnetized accretion disc, we expect to have hot electrons primarily concentrated in the coronal and outflowing regions characterized by $\beta \lesssim 1$.  Note that although the quantitative formula we use is only strictly valid for heating due to dissipation at the smallest scales of the MHD turbulent cascade and not magnetic reconnection, heating due to the latter has a qualitatively similar dependence on $\beta$ \citep{Numata2014}. 
To calculate the total heating rate per unit volume, $Q \equiv \rho T_g u^\mu \partial_\mu s_g$, we use the model detailed and tested in \citet{Ressler2015}, which self-consistently captures the numerical heating provided by the ideal conservative GRMHD evolution. \citet{Ressler2015} show that this method accurately calculates the heating rate in several test problems, including strong shocks and forced MHD turbulence.  Finally, we evolve the conductive flux identically to \citet{Ressler2015} using the model of \citet{ManiModel},  where we parametrize the electron thermal conductivity with a dimensionless number $\alpha_e$, related to the conductivity, $\chi_e$ via 
\begin{equation}
  \chi_e = \alpha_e c r.
  \end{equation}
  For the present work, we focus on $\alpha_e = 10$ and $\alpha_e =0$. The former essentially saturates the heat flux at its maximum value of $u_{e} v_{t,e}$, where $u_e$ is the electron internal energy per unit volume and $v_{t,e}$ is the electron thermal speed, while $\alpha_e = 0$ corresponds to zero heat flux.

The only free parameter in our electron model is the dimensionless electron conductivity, $\alpha_e$, since we have fixed the electron heating model as described above.  Note that there are, however, significant uncertainties introduced by the uncertainty in the poorly constrained macroscopic parameters of the system (e.g., magnetic flux and black hole spin).

\section{Radiation Transport}
\label{sec:rad}
To calculate model spectral energy distributions (SEDs), we use the Monte Carlo radiation code {\tt GRMONTY} \citep{Dolence2009} adapted to use our evolved electron temperature to calculate the emissivity and scattering/absorption cross-sections.  We include synchrotron emission/absorption and inverse Compton scattering.  The emission is calculated in post-processing and does not affect the flow dynamics. Furthermore, we also generate radio and infrared images using the ray-tracing code {\tt iBOTHROS} \citep{Noble2007} which includes synchrotron emission/absorption.  

When calculating the spectrum, we average the emission over azimuthal observing angle in order to reduce noise.  This does not qualitatively affect the time-averaged spectrum and only very modestly reduces temporal variability (as we have determined using a subset of the simulation outputs).  Furthermore, we also make the ``fast light'' approximation, meaning that we compute a single spectrum by propagating photons on a fixed time slice of fluid quantities.  This amounts to assuming that the light propagation time across the domain is small compared to the dynamical time and should not be a dominant source of error. 

While the GRMHD simulation is scale free, the radiation transport depends on the physical mass scale of the accretion disc.  This dependence can be represented by a single free parameter, namely, the mass unit, $M_{\rm unit}$, which is a number in grams that converts the simulation density to a physical density (and thereby fixes the physical accretion rate).  We set this free parameter by normalizing the time-averaged flux at 230 GHz to the observational value of $2.4$ Jy \citep{Doeleman2008}.

Finally, in order to limit the emission to regions of the simulation in which we can reasonably trust the fluid thermodynamics, we impose a limit on the flow magnetization $\sigma = b^2/\rho c^2$.  That is, we only consider emission that originates or scatters from regions of $\sigma < 1$. The thermodynamics in regions with larger $\sigma$ become uncertain in conservative codes because small errors in the total energy (which is dominated by magnetic energy) lead to large errors in the internal energy. Note that this is true for both the underlying GRMHD entropy and temperature and not just the electron temperature.  The motivation for our particular maximum value of $\sigma$ and the effects of varying this parameter are described in Appendix \ref{App:sigma}.

\section{Accretion Disc Model}
\label{sec:fid}

We initialise the simulation with the now ``standard'' \citet{Fishbone1976} equilibrium torus solution with a dimensionless spin, $a = 0.5$, inner radius, $r_{\rm in} = 6 r_g$, and with the maximum density of the disc occurring at $r_{\rm max} = 13 r_g$ (see Appendix~\ref{App:coords} for more details). Here $r_g = GM/c^2$ is the black hole gravitational radius. The adiabatic index of the gas is taken to be $\gamma = 5/3$, appropriate for ions with sub-relativistic temperatures.  We initialize the torus with a single magnetic field loop in the $(r,\theta)$ plane, as we discuss in Appendix~\ref{App:coords}.

We take the adiabatic index of the electron fluid to be $\gamma_e = 4/3$, appropriate for relativistically hot electrons, and initially set $u_e = 0.1 u_g$ and the electron heat flux to zero.   We apply the floors on internal energy (both for the total gas and electrons) and density in the drift frame of the plasma as described in Appendix \ref{App:coords}. Here we use the same floor prescription for the electrons as in \citet{Ressler2015}.  We run the simulation for a time of $19,000$ $M$, which is long enough for the inner $r \lesssim 25 r_g $ portion of the disc to be in inflow equilibrium.  The outflow, however, travels at higher velocities so that it takes $\lesssim 1000$ $M$ for the flow to reach $100 r_g$.

Note that we have assumed a constant value for both the electron and total gas adiabatic indices.  \citet{Sadowski2016} implemented temperature-dependent adiabatic indexes and showed that while $\gamma_e$ was always $\approx 4/3$ in the domain of interest, the total adiabatic index varied from $5/3$ in the midplane to $4/3$ in the polar regions, meaning that assuming $\gamma=5/3$ (as we do here) overestimates the gas temperature by about a factor of 2. However, their resulting electron temperatures were qualitatively very similar to those in the constant adiabatic index model (see their Figure 5), so we do not expect this approximation to have a significant effect on our results.

Figure~\ref{fig:ic_grid} shows our computational grid, which is uniformly discretized in ``cylindrified'' and ``hyper-exponential'' modified Kerr-Schild (MKS) coordinates as described in Appendix \ref{App:coords}. The grid extends from an inner radius of $r_{\rm in} = 0.8 (1 + \sqrt{1-a^2})$ $r_g$ ($\approx 1.62 r_g$ for $a=0.5$) to an outer radius of $r_{\rm out} = 10^5 r_g$. In contrast to the cylindrified and hyper-exponentiated coordinates we use in {\tt HARMPI}, we use standard MKS coordinates in {\tt iBOTHROS} and {\tt GRMONTY}.  As photons propagate between grid points, the radiation transport algorithms require frequent evaluation of the connection coefficients which are analytic in MKS but require multiple numerical derivatives in the cylindrified coordinates.  The latter greatly increases the computational cost of these methods, which is not an issue for {\tt HARMPI} because after it evaluates the connection values once at the beginning of the simulation at each grid point, it stores them for future use. To read in data from {\tt HARMPI}, we first use the Jacobian of the coordinate transformation to convert all 4-vectors and then interpolate onto the grids of {\tt iBOTHROS} and {\tt GRMONTY}. 

\begin{figure}
\includegraphics[scale = .1]{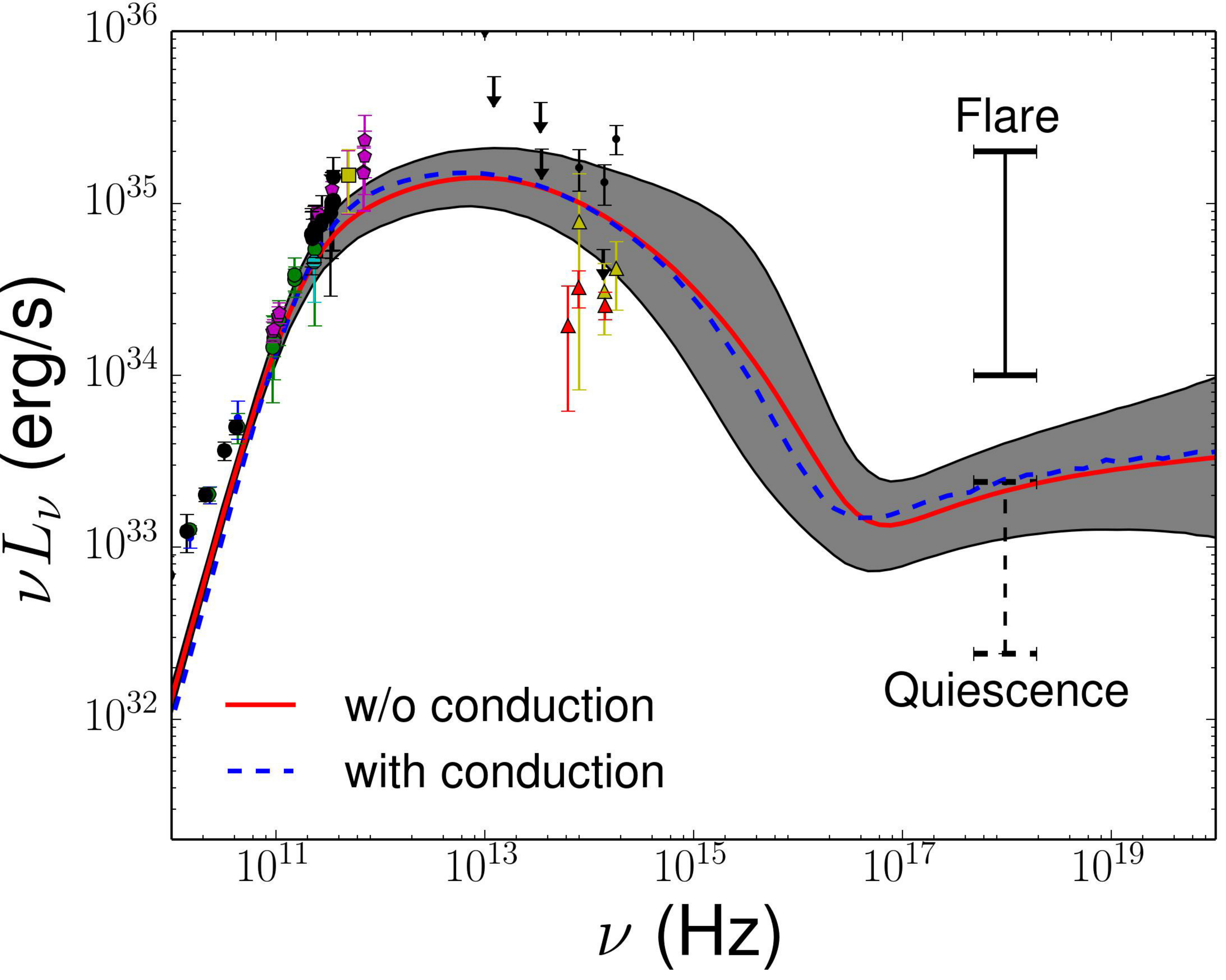}
\caption{Spectral Energy Distribution (SED) for our fiducial model averaged over $15,000 - 19,000$ $r_g/c$ (about 1 day for Sgr A*), as observed at an inclination angle of $45^\circ$ with respect to the spin axis of the black hole.  We show results with and without anisotropic electron conduction (with the dashed blue and solid red lines, respectively). The shaded grey region represents the $1 \sigma$ time-variability of the SED over this time interval without conduction (the time variability with conduction is indistinguishable, so it is not plotted here). Data points represent various observations and upper limits (see Appendix \ref{sec:obs}). The solid vertical line in the X-rays roughly represents the range of observed flares in Sgr A* \citep{Neilsen2013}, while the dashed vertical line represents ``quiescent'' emission (i.e. between 10 -100\% of the total total quiescent emission observed from Sgr A*; \citealt{Baganoff2003} ).  The SED is normalized to match the observed 230 GHz flux. }
\label{fig:SANEspec}
\end{figure}

\section{Results}
\label{sec:res}
\subsection{Basic Flow Properties}
Figure \ref{fig:SANEFluid} shows the time and azimuthally averaged electron temperature, density, and heat flux relative to the field-free value, as well as 1D angular profiles of the plasma beta parameter, $\beta$, magnetization, $\sigma = b^2/\rho$, electron heating fraction, $f_e$, and dimensionless electron temperature, $\Theta_e\equiv k_B T_e/ m_e c^2$. The 1D profiles are additionally averaged over radius from the horizon to $25 r_g$.  Our 3D simulations reproduce the general qualitative result of \citet{Ressler2015}'s 2D simulation: the hottest electrons are concentrated in the lower density coronal and funnel wall regions, while the midplane of the disc remains relatively cold.  Furthermore, we find that the anisotropic heat flux is suppressed by a factor of $\sim 5-10$ relative to the isotropic heat flux, roughly equivalent to the 2D result. This is because the magnetic field is, on average, primarily toroidal, while the temperature gradients are primarily poloidal.  We note that the total heating rate integrated over the volume with $\sigma <1$ between the event horizon and the inflow equilibrium radius\footnote{The integrated heating rate is calculated as $\int\limits_{r_H}^{25 r_g} -Q u_t \sqrt{-g} dx^1 dx^2 dx^3$, where $r_H$ is the radius of the event horizon. } ($\sim 25 r_g$) is $\sim 0.39 \%$ of $|\dot M c^2|$, well below the efficiency predicted by the \citet{Novikov1973} (NT) model for a disc extending out to $25 r_g$ with a spin of $a = 0.5$ ($6.3\%$). This result is not necessarily surprising; the thin disc efficiency assumes that all of the gravitational binding energy of the disc must be dissipated and radiated away and that outflow is negligible.  RIAF discs, on the other hand, are typically characterized by significant outflow in the form of Poynting and turbulent energy flux so that the energy going into dissipation can be much less (though the latter is typically only a small fraction of $\dot M c^2$, e.g. \citealt{Yuan2012}, the former can be significant, e.g. \citealt{McKinney2004}). However, we find that our calculation of the total heating rate has significant contribution from the negative heating in the polar regions (discussed in \S \ref{sec:Thermpol}) which is a consequence of numerical diffusion.  If we focus exclusively on the disc, excluding negative heating rates in the polar regions, the heating rate in the same volume totals $\sim 4.6 \%$ of $|\dot M c^2|$, much higher, of order the NT efficiency.

The simulation has a significant amount of magnetic flux threading the black hole, with a time averaged value of $\Phi_{BH} \approx 40$ $(\dot M c )^{1/2} r_g$, which can be compared to the typical saturation value of a Magnetically Arrested Disc, $\approx 50$ $(\dot M c )^{1/2} r_g$ \citep[MAD,][]{2003PASJ...55L..69N,Sasha2011}, at which the excess flux impedes the inflowing matter. Interestingly, in a few test runs varying the magnetic flux, we have found that as long as the magnetic flux is below this saturation value, the qualitative features of the spectrum are not strongly dependent on the flux threading the black hole.  Note that this is true only when normalising the spectrum to the 230 GHz flux by varying the accretion rate. Higher (lower) magnetic flux values tend to require smaller (higher) accretion rates.  If instead we increased $\Phi$ at a fixed accretion rate we would expect significant differences in the spectrum (e.g., higher flux, higher peak frequency, etc).

\subsection{Spectra and Images}
Figure \ref{fig:SANEspec} shows the SED of our model averaged from $15,000-19,000$ $r_g/c$ (a time of about 1 day for Sgr A*), at an inclination angle of $45^{\circ}$ with and without thermal conduction and with time-variability shown by the shaded region. To normalize the $230$ GHz flux, the simulation required a time-averaged accretion rate of $1.1 \times 10^{-8} M_{\odot}\, {\rm yr}^{-1}$, or $\sim$ $1.2 \times 10^{-7}$ $\dot M_{\rm Edd}$ for Sgr A*. This is in reasonable agreement with the estimate of $6 \times 10^{-8} M_{\odot} \, {\rm yr}^{-1}$ provided by the inflow-outflow model of \citet{Shcherbakov2010} and falls within the constraints set by radio polarization measurements \citep{Marrone2007}.  Interestingly, this accretion rate is about two orders of magnitude less than the accretion rate at the Bondi radius inferred from X-ray observations \citep{Baganoff2003}, suggesting the existence of a strong, large scale outflow.

\begin{figure*}
\includegraphics[scale = .1]{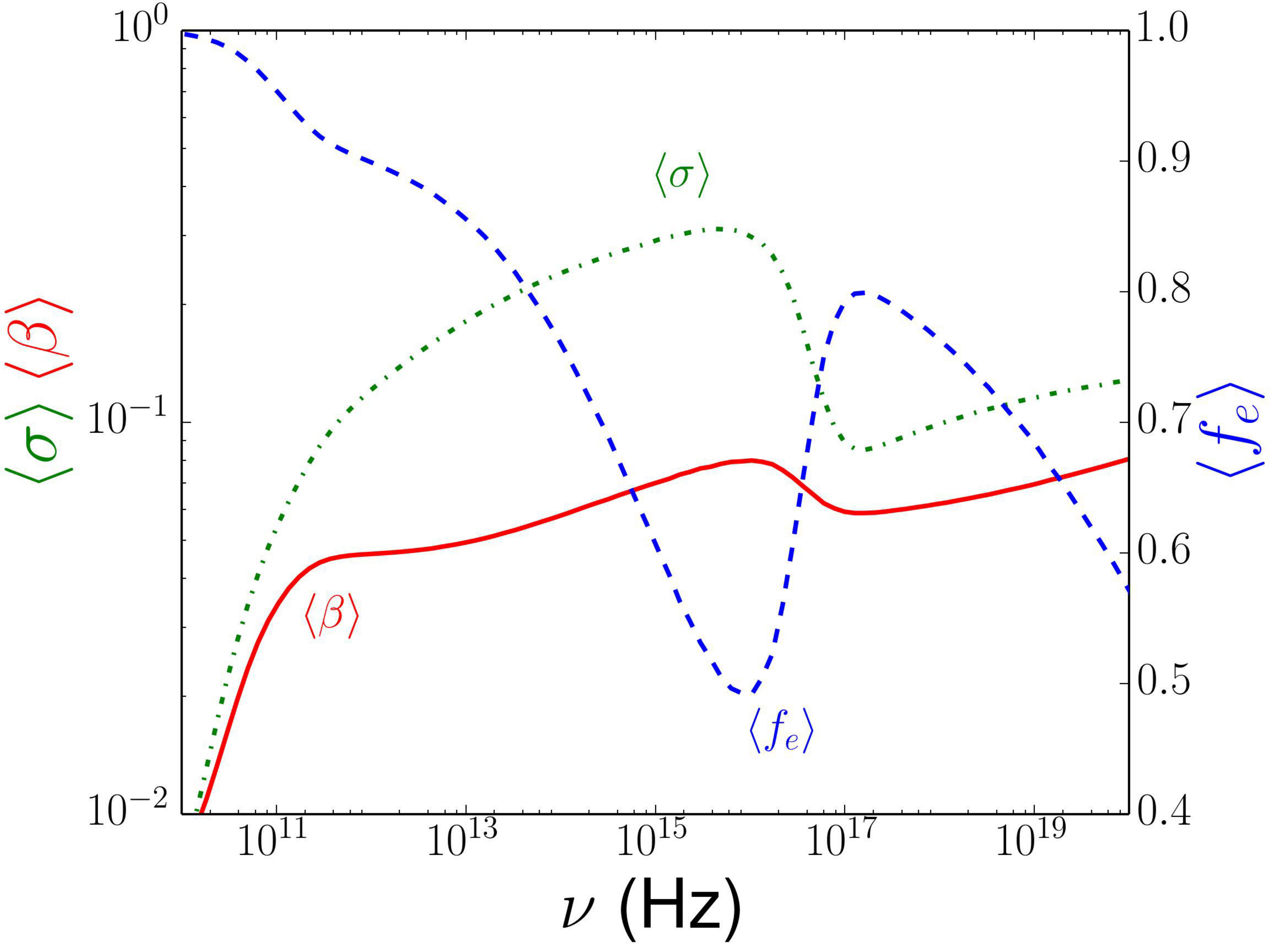} {\color{white} space}
\includegraphics[scale = .1]{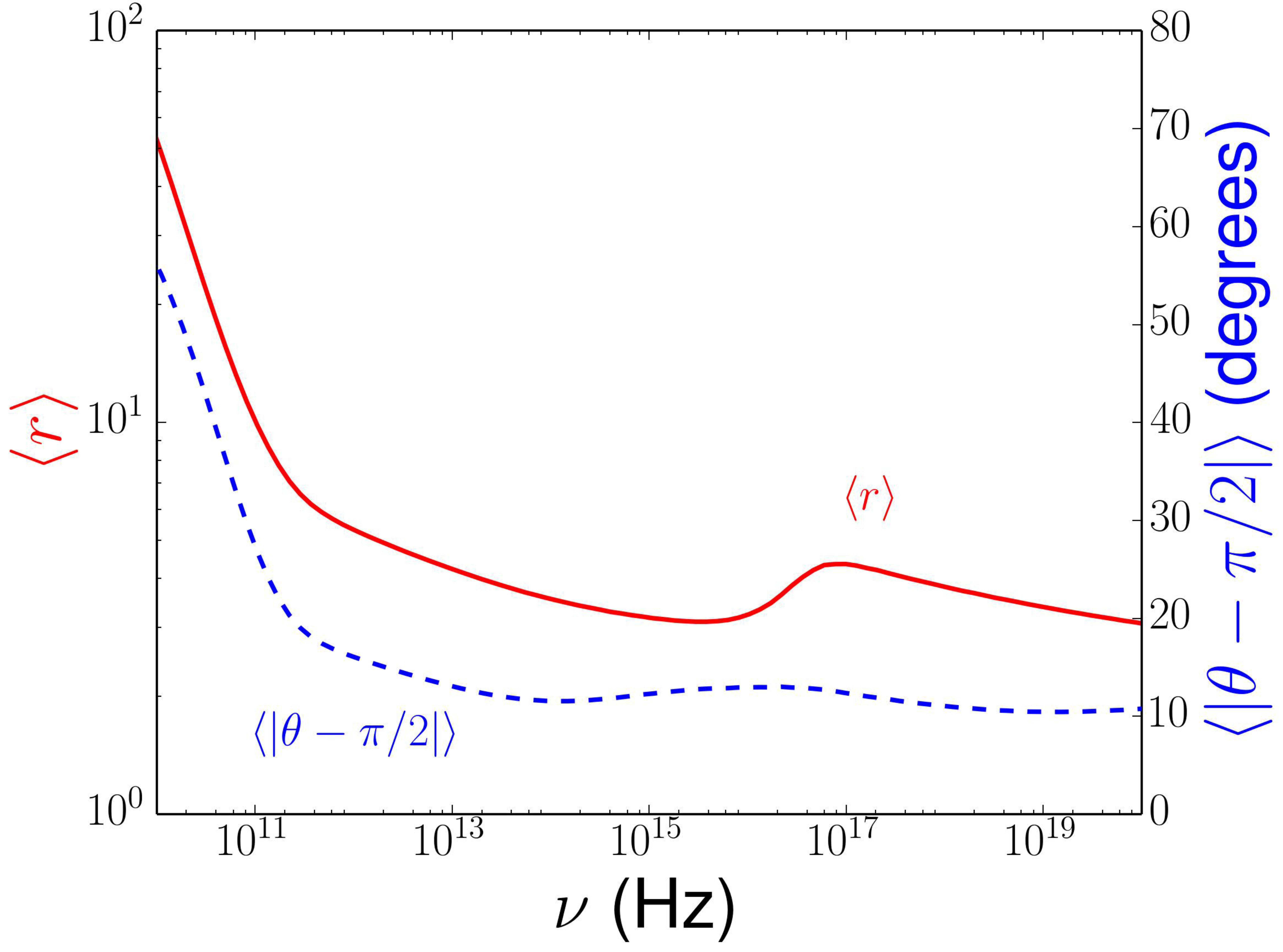} 
\includegraphics[scale = .1]{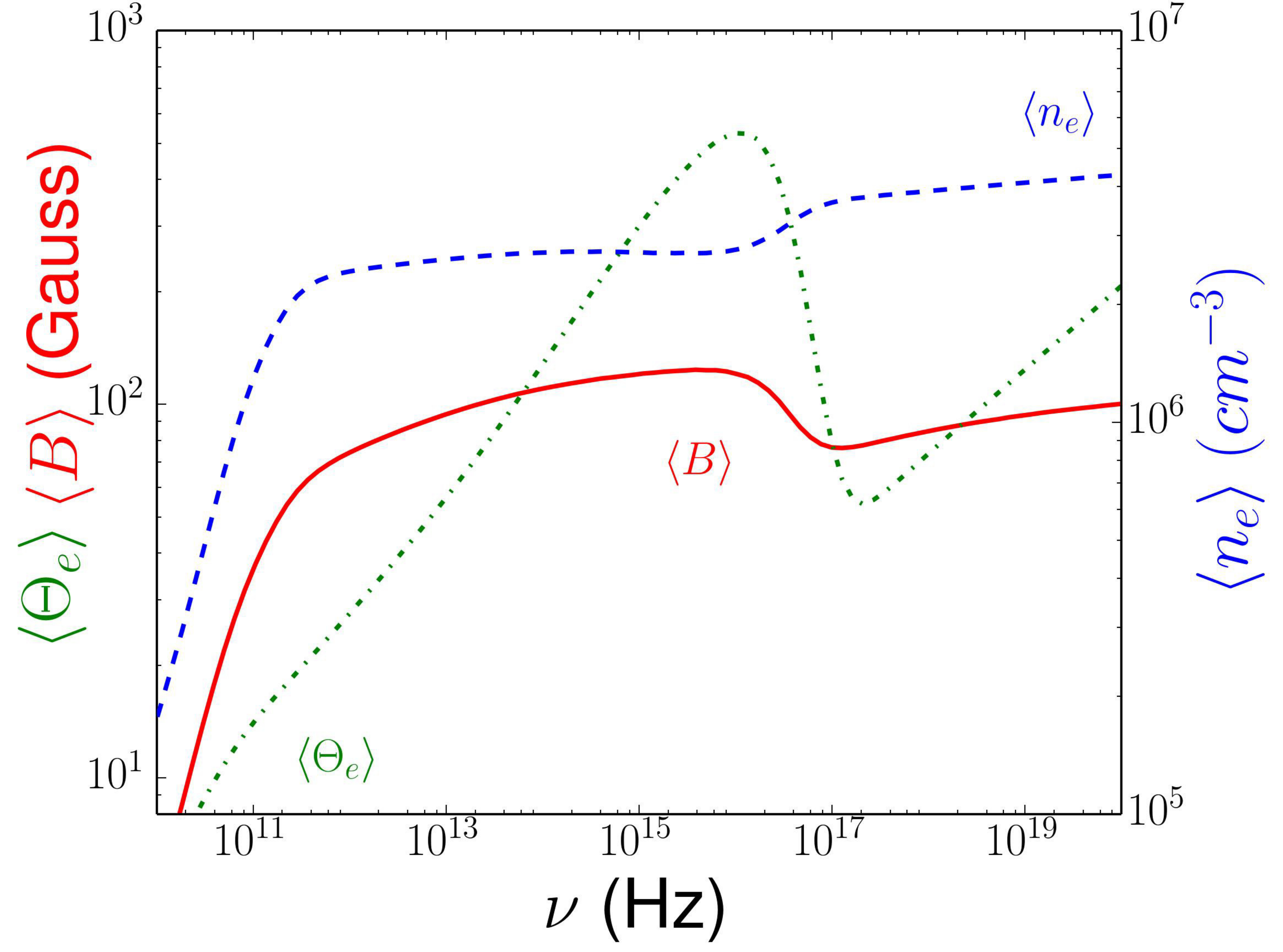}{\color{white} space}
\includegraphics[scale = .1]{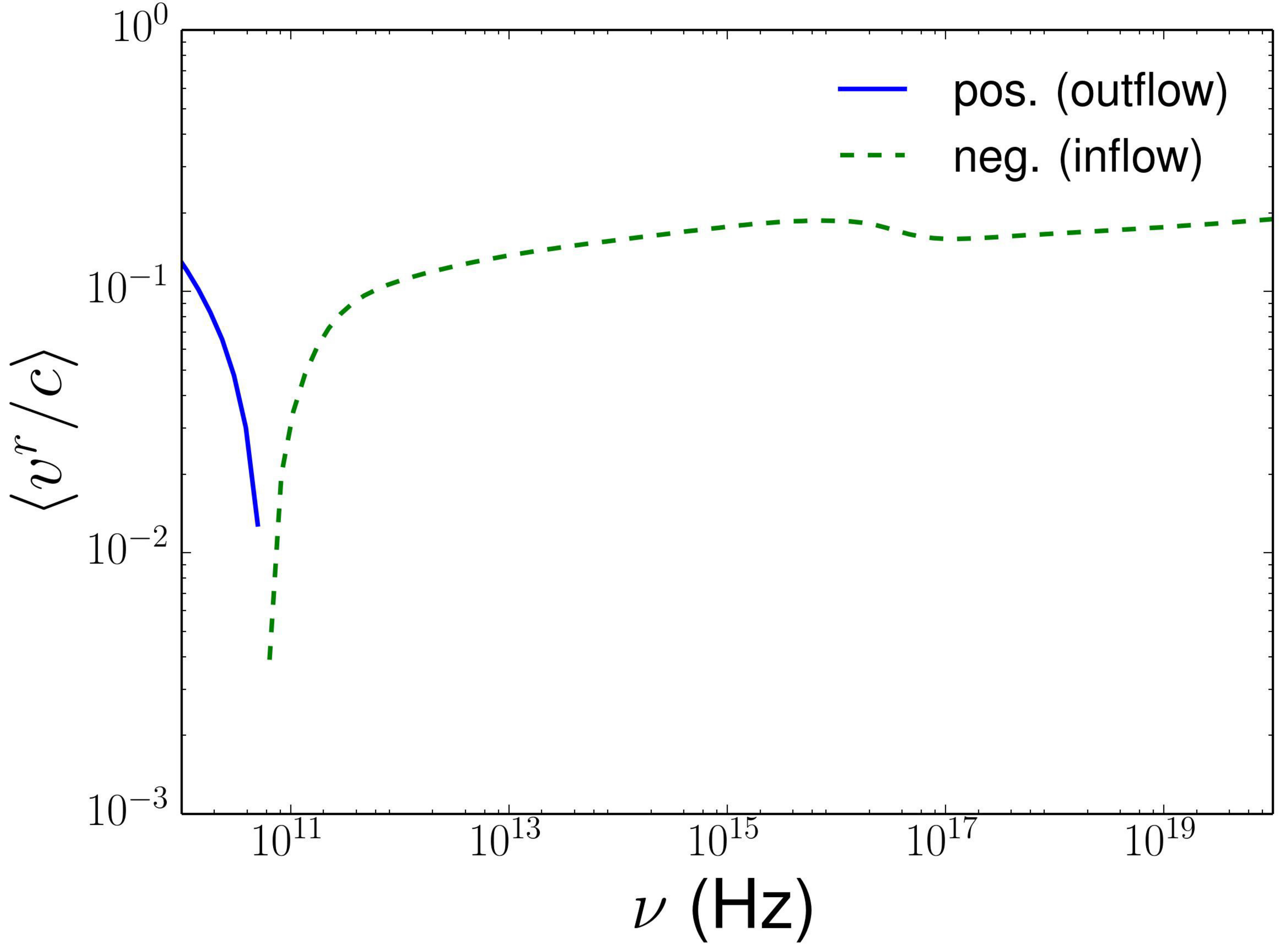}

\caption{Fluid quantities at the point of origin averaged over individual photons as a function of observed photon frequency for the spectrum shown in Figure \ref{fig:SANEspec}.  Plotted are the magnetization, $\langle \sigma \rangle  = \langle b^2\rangle / \langle \rho\rangle$, the plasma $\langle \beta \rangle = \langle P_g\rangle / \langle P_m\rangle$, the electron heating fraction, $\langle f_e \rangle$, the Boyer-Lindquist (BL) radial coordinate, $\langle r \rangle$, the deviation of the BL polar angle from the midplane, $\langle |\theta - \pi/2|\rangle $, the magnitude of the fluid-frame magnetic field, $\langle B\rangle$ = $\langle \sqrt{b^\mu b_\mu} \rangle $, the electron number density, $\langle n_e\rangle$, and the dimensionless electron temperature in units of the electron rest mass, $\langle \Theta_e\rangle = \langle k_B T_e/ m_e c^2\rangle$,  as well as the radial velocity, $v^r  = \sqrt{g_{11}} \langle u^{x1}/u^t \rangle $.   The optically thick low frequency synchrotron emission (below $\sim 230$ GHz) comes from larger radii in the outflow away from the midplane where $\beta$ is smallest and hence the electron heating fraction, $f_e$, is largest.  It is interesting to note that despite the larger $f_e$ in these regions that the electron temperatures are quite modest $(\Theta_e \lesssim 10)$ (see Section \ref{sec:lowfreqrad} for more details). The higher frequency emission (above $\sim 230$ GHz) is emitted and/or scattered from smaller radii close to both the horizon and the midplane.  In these regions, the temperatures are much larger (reaching $\Theta_e \sim 100$ and above), due to the increased turbulent heating in the disc fuelled by the deeper gravitational potential.   Between $230$ GHz and about $10^{17}$ Hz the emission is predominantly optically thin synchrotron, while above $10^{17}$ Hz the emission is predominantly inverse Compton scattering of IR photons by IR-emitting electrons.  This explains the transitions in the average fluid quantities at about $10^{17}$ Hz.}
\label{fig:freqplots}
\end{figure*}

It is convenient to interpret the spectrum using the luminosity-weighted fluid quantities at the last scattering surface (this is simply the location of the emitting regions for photons optically thin to scattering).  These are shown as a function of frequency in Figure \ref{fig:freqplots}.  The spectrum can be decomposed into three distinct regions:
\begin{enumerate}
\item Below about $\sim 230$ GHz the emission is optically thick synchrotron and originates at larger radii ($\sim 10-200 r_g$) in the outflow ($v^r \sim 0.01 - 0.1 c$) of the corona/funnel ($|\theta-\pi/2| \sim 20^\circ - 60^\circ$).
\item Between $\sim 230$ GHz and $\sim 10^{17}$ Hz $\simeq 0.5$ keV the emission is optically thin synchrotron from radii close to the horizon ($\lesssim 10 r_g$) and closer to the midplane ($|\theta-\pi/2| \sim  10^\circ - 30^\circ)$. On average, the emitting regions are inflowing.
\item Above $\sim 10^{17}$ Hz $\simeq 0.5$ keV the luminosity-weighted number of scatterings sharply transitions from $\sim0$ to $\sim1$, indicating that the X-ray emission is dominated by inverse Compton scattering.  More precisely, by computing the luminosity-weighted photon energy gain per scattering and the luminosity-weighted pre-scattering frequency, we find that the X-ray emission is dominated by infrared photons ($\sim 10^{13} -10^{15}$ Hz) scattered by electrons emitting synchrotron radiation in the IR.  The latter point can also be seen in the correspondence between the luminosity-weighted fluid quantities at the point of origin for the IR and X-ray photons in Figure \ref{fig:freqplots}.
\end{enumerate}

Figure \ref{fig:SANEspec} shows that the spectra of models with and without anisotropic electron thermal conduction are nearly indistinguishable from each other.  We have found this to be a robust result for the Standard and Normal Evolution (SANE) accretion flows \citep{2012MNRAS.426.3241N} without dynamically-important magnetic flux over a wide variety of initial conditions, black hole spin, and magnetic flux (that is, for fluxes less than the MAD saturation limit).  

We find that our model produces significant X-ray and NIR variability, which qualitatively agrees with the observed flaring behaviour of Sgr A* (see \S \ref{sec:tvar}), though for this particular model we do not see strong X-ray flaring events ($\gtrsim 10$ times quiescence) and the quiescent X-ray flux may be moderately overpredicted.  We also find that our fiducial model has a spectral slope near $230$ GHz that agrees well with observations, but at $\lesssim 10^{11}$ Hz the slope becomes steeper than that observed, $d \log(F_\nu)/d\log\nu \approx 0 $, resulting in an underprediction of the low frequency emission  (see \S~\ref{sec:lowfreqrad} for more details).  

\begin{figure}
\includegraphics[scale = .12]{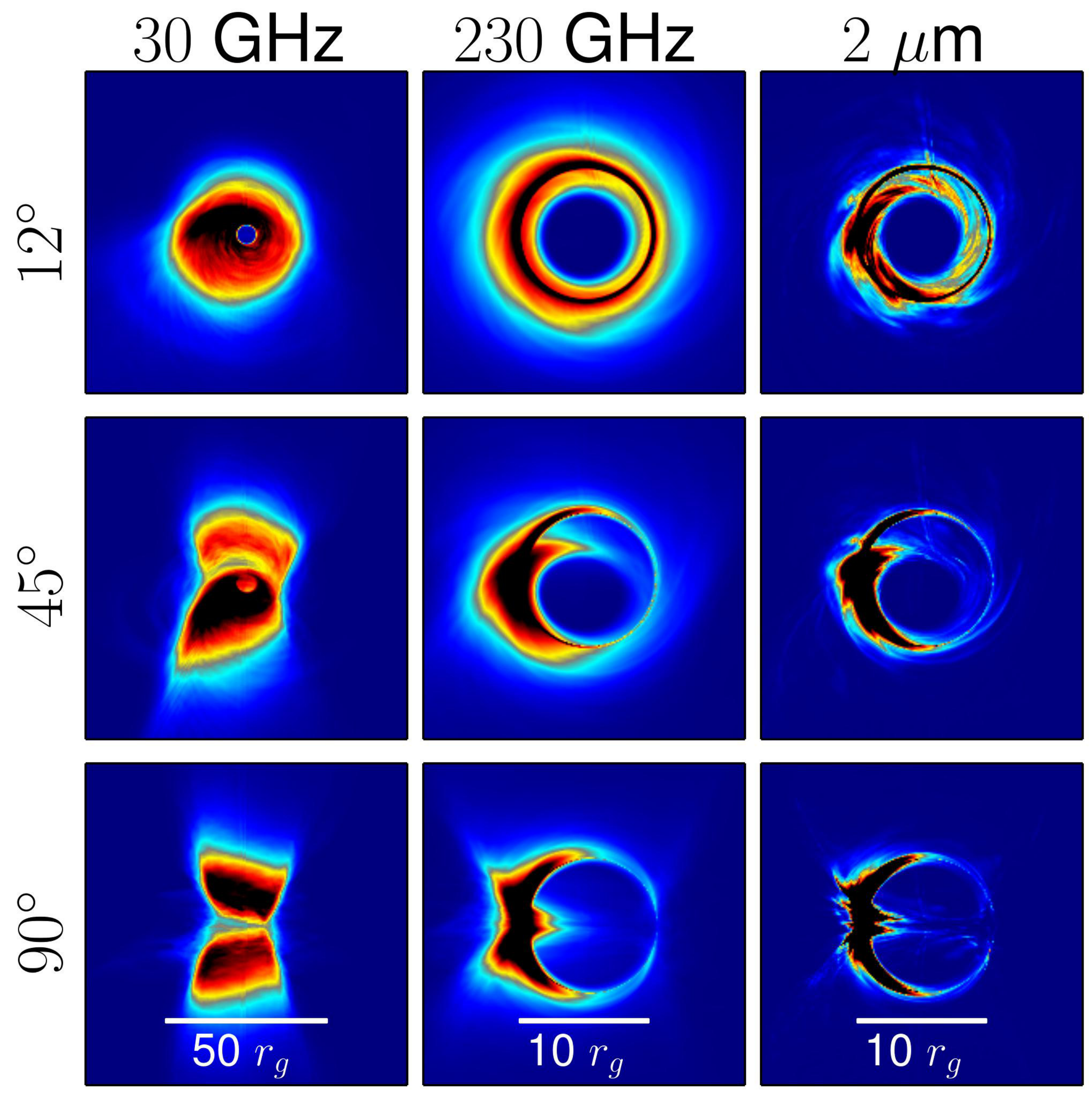}
\caption{Linear intensity maps for our fiducial model of Sgr A* without electron thermal conduction (the effect of conduction on the images is negligible) at 30 GHz (left column), 230 GHz (middle column), and 2 $\mu$m (right column) for inclination angles of $12^\circ$ (top row), $45^\circ$ (middle row) and $90^\circ$ (bottom row). The inclination of $90^\circ$ is edge-on while $12^\circ$ is nearly face-on.  Images are averaged over time from $15,000 - 19,000$ $r_g/c$ and normalized such that the $230$ GHz flux is $2.4$ Jy \citep{Doeleman2009}.  The physical size of the 30 GHz images is $100$ $r_g$ $\times$ 100 $r_g$ while the physical size of the $230$ GHz and 2 $\mu$m images is $25$ $r_g$ $\times$ 25 $r_g$. For all inclination angles, the photon ring is clearly visible at 230 GHz, the frequency at which the Event Horizon Telescope will be able to spatially resolve Sgr A* \citep{Doeleman2008}, and is also clearly visible at 2 $\mu$m, the wavelength of interest to GRAVITY \citep{Gillessen2010}. 
The low frequency radio emission is dominated by the outflow at large radii, consistent with previous phenomenological models of Sgr A* (e.g. \citealt{Falcke1995, Yuan2002, Moscijetorig,Mosci2014,CK2015}).} 
\label{fig:SANEimage}
\end{figure}

Figure \ref{fig:SANEimage} shows time-averaged 30 GHz, 230 GHz, and 2 $\mu$m images without electron thermal conduction at $12^{\circ}$, $45^\circ$ and $90^\circ$ (images with electron conduction look nearly identical). In generating these images we used {\tt iBOTHROS} and neglected Inverse Compton scattering, as appropriate for such low frequencies. The photon ring is clearly visible at both 230 GHz and 2 $\mu$m.  This bright circle of emission surrounding the shadow of the black hole is the observational signature of the effects of the circular photon orbit and strong lensing on the small emitting region in the simulations.  A primary goal of the Event Horizon Telescope is to measure the size of this ring in order to probe the strong field limit of general relativity.  The lower frequency images are dominated by disc-jet emission from larger radii while the higher frequency emission is dominated by disc emission close to the black hole.   

An important property of our results is that they self-consistently produce the ``disc-jet'' structure appealed to in previous phenomenological models (e.g.; \citealt{Falcke1995,Yuan2002,Moscijetorig,Mosci2014,CK2015}).   This is clear in the luminosity-weighted fluid quantities in Figure \ref{fig:freqplots} which show a transition from emission dominated by inflowing equatorial material above $10^{11}$ Hz to emission dominated by outflowing polar material at lower frequency.    The disc-jet structure is particularly clear in the images in Figure \ref{fig:SANEimage}, which show that the outflow dominates at lower frequency while the very compact emitting disc dominates at higher frequencies.  This type of structure naturally occurs in our model because of the strong $\beta$ dependence of the electron heating fraction, $f_e$, which suppresses electron heating in the midplane in favour of the polar regions.

\subsection{Time Variability}
\label{sec:tvar}

\begin{figure*}
\includegraphics[scale = .1]{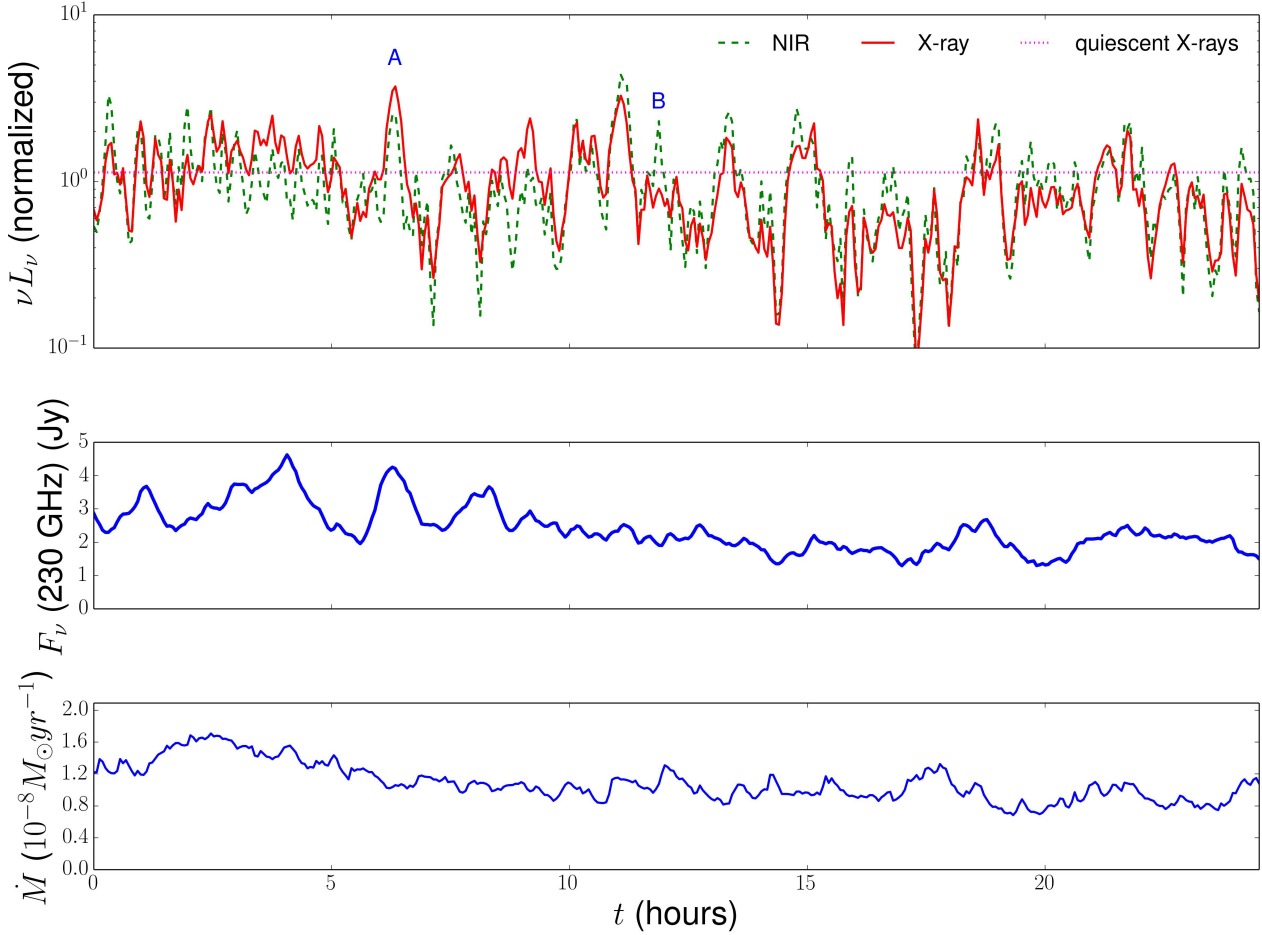}\\

\caption{\emph{Top Panel}: $\nu L_\nu$ as a function of time at an inclination angle of 45$^\circ$ for NIR (2.18 microns) and X-ray (2-8 keV) bands during the time interval $15,000-19,000$ $r_g/c$ or $\sim 1$ day for Sgr A*.  Each light-curve is normalized to the mean value of $\nu L_\nu$ over the entire interval. Also plotted is the total quiescent X-ray flux as observed by \emph{Chandra} \citep{Neilsen2013}, with the same normalization as our model's X-ray light curve. Though only $\sim 10 \%$ of this quiescent emission is believed to originate from the inner accretion flow, the total quiescent flux is the relevant threshold for X-ray flares.  Both the X-ray and NIR emission show order of magnitude variability on time-scales of $\sim 0.5$ hours and are strongly correlated.  Note that, as observed in Sgr A*, each X-ray flare is associated with a NIR flare (e.g. the peak labeled ``A''), but there are also a few candidates for NIR flares without X-ray counterparts (e.g. the peak labeled ``B''), which are also observed in Sgr A*. \emph{Middle Panel}: 230 GHz flux in Jy as a function of time at an inclination angle of 45$^\circ$.  Compared to the NIR and X-ray flux, the mm flux varies only weakly over this time interval (note the linear scale), consistent with observations.  \emph{Bottom Panel}: Accretion rate in units of $10^{-8}$ solar masses per year. The variability in $\dot M$ has roughly the same time scale as the variability in NIR and X-ray emission because the latter originate close to the horizon. However, in detail the emission is not well-correlated with fluctuations in $\dot M$. }
\label{fig:Fvt}
\end{figure*}

Figure \ref{fig:Fvt} shows the light curves of our fiducial model in the mm, NIR, and X-ray frequencies compared to the accretion rate of the disc over the same time interval.  Our simulations show significant and correlated time variability in the NIR and X-ray bands while the mm emission is significantly less variable.  

The general correspondence between the NIR and X-ray light curves is due to the fact that the flares are caused by localized hot spots of low $\beta$ that emit high levels of NIR synchrotron emission, a fraction of which is then additionally upscattered to the X-rays.  Each large spike in X-rays is accompanied by a comparably large spike in the NIR emission (e.g. the X-ray peak labeled ``A''), in agreement with observations of Sgr A* that find that all X-ray flares have NIR counterparts (see, e.g., Table 3 in \citealt{Eckart2012} for a recent summary).  On the other hand, there are a few large spikes in the NIR emission that are accompanied by only relatively small increase in the X-ray flux (e.g. the NIR peak labeled ``B''). 
%Though technically the X-ray emission at these instances is above the quiescent upper limit, 
Since we have not tuned our model to precisely match the time-averaged quiescent X-ray flux, the key feature here is the significant increase in IR luminosity without a corresponding increase in X-ray luminosity (and not necessarily whether the X-ray luminosity is above the quiescent threshold).
%Though technically the X-ray emission at these instances is above the quiescent upper limit, the definition of a ``flare'' is not rigorously defined since we have not tuned our model to precisely match the time-averaged quiescent X-ray flux. 
Therefore, these particular NIR flares are consistent with lacking a strong X-ray counterpart, which are are also observed in Sgr A* (e.g., \citealt{Hornstein2007}, \citealt{Trap2011}).  
Furthermore, we find that the X-ray emission in our model in fact has no well defined ``quiescent state'' but is rather constantly flaring.  This has been suggested for Sgr A*, where the observed ``quiescent state'' could be a collection of undetectable flares \citep{Neilsen2013}.

The X-ray ``flares'' in our model shown in Figure \ref{fig:Fvt} are relatively weak in magnitude, with luminosities peaking at factors of only a few times the quiescent level. During this time interval, we find no evidence for the strong X-ray flares observed in Sgr A* which range from $\sim$ 10 to $\gtrsim 100$ times the quiescent level (e.g.; \citealt{Trap2011} and references therein; \citealt{Neilsen2013}). While this could point to the need for nonthermal particles (see, e.g. \citealt{Ball2016}), it could also be a consequence of the limited time interval considered here ($\sim$ one day, while Sgr A* has major flares only $\sim$ once per day) or the particular parameters used in our model (e.g., spin and magnetic flux). This will be investigated in future work.

\subsection{Dependence of Observables on Disc Parameters}
Here we briefly describe the qualitative effects of varying several parameters of our fiducial model. These effects are not unique to our electron model but are more general properties of radiative GRMHD models of low $\dot M$ discs as seen in previous parameter studies (e.g. \citealt{Mosci2009, CK2015})

\emph{Spin}: Larger black hole spin tends to increase both NIR and X-ray emission. This is because high spin black holes have inner-most stable circular orbits (ISCOs) that are closer to the event horizon, which means that the accretion disc will extend to smaller radii where the temperatures are generally higher due to the deeper gravitational potential.  These higher temperature regions emit at higher frequencies.

\emph{Inclination Angle}: Inclination angles closer to 90$^\circ$ (edge-on), tend to have larger NIR and X-ray emission (relative to the mm emission) than inclination angles closer to 0$^\circ$ (face-on).  This is primarily due to Doppler boosting caused by rotation of the disc.  When looking edge-on, Doppler beaming leads to the observed emission being dominated from the side of the disc that is moving towards us, which will also be Doppler blue-shifted from the fluid frame frequency.  This increases the relative NIR and X-ray emission compared to the mm emission.  On the other hand, when looking face-on, the motion of the disc is perpendicular to the line of sight and Doppler effects are minimized (though there can still be Doppler effects from the outflow).

\subsection{Convergence of Spectra}
\label{sec:therm}

\begin{figure}
\includegraphics[scale = .105]{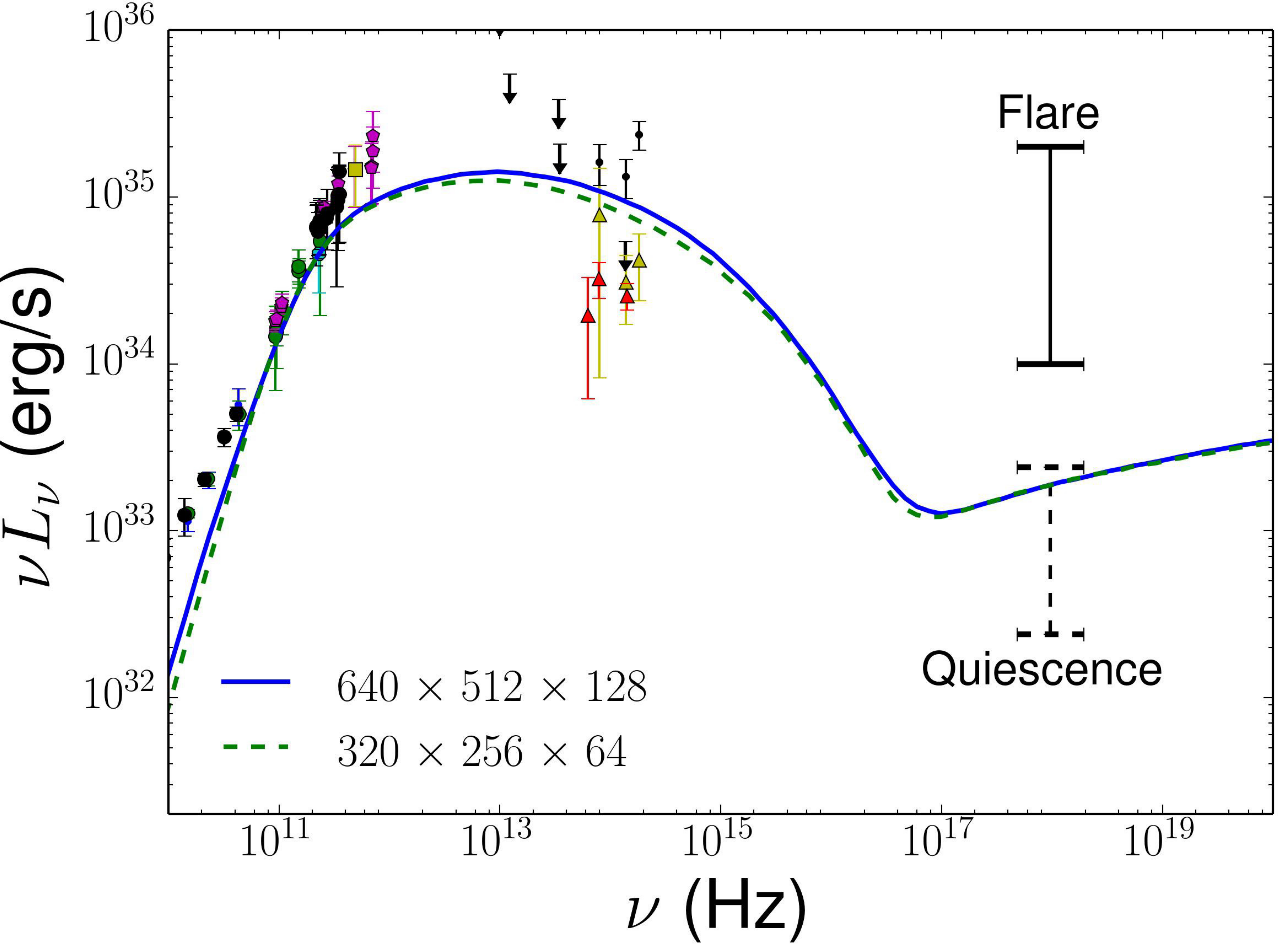}
\caption{Spectral Energy Distribution (SED) for our fiducial model averaged over $11,800 - 12,300$ $r_g/c$ at an inclination angle of $45^\circ$ at two different resolutions (please see the legend). The higher resolution simulation was initialized with the results of the low resolution simulation at $10,000$ $r_g/c$ and run for an additional $2,300$ $r_g/c$.   The spectra only display minor differences below $\sim 10^{11}$ Hz and above $\sim 10^{14}$ Hz. }
\label{fig:spec_convergence}
\end{figure}

In order to test whether our results are converged, we restarted the fiducial model described in Section~\ref{sec:fid} at double the resolution in each direction (namely, $640 \times 512 \times 128$).  We did this by copying the fluid quantities at a particular time in each cell on the lower resolution grid into 8 cells on the higher resolution grid and using this as an initial condition. To prevent magnetic monopoles from being generated by the numerical interpolation, we operated on the magnetic vector potential instead of the magnetic field directly.  We then ran for 2000 M, computed the time-averaged spectra, and compared to the lower resolution spectra that had also run for an additional 2000 M.  Since 2000 M is roughly enough time for the inner $\sim 15 r_g$ of the disc to accrete and for the outflow to reach beyond $100 r_g$, we are reasonably confident that the simulation has had enough time to evolve dynamically from the initial restart. Figure \ref{fig:spec_convergence} shows that the time averaged spectra are qualitatively the same. Quantitatively, the differences are minor, though interestingly the low frequency slope in the higher resolution simulation is slightly closer to observations.  Unfortunately, doubling the resolution even further to see if this trend continues is too computationally expensive with our current resources; in fact, even running the $640 \times 512 \times 128$ simulation for much longer is pushing the limit of what we can afford for the present work. With that said, the fact that the differences between the spectra at these two different resolutions are almost negligible is encouraging and provides some assurance that the observational features of our model are not strongly dependent on resolution.

\section{Thermodynamics in the Polar Outflow}
\label{sec:Thermpol}
The low frequency radio emission in our simulation generally originates within or near to what has in past work been described as the jet ``sheath'' (e.g. \citealt{Moscijetorig}), which is the portion of the outflow that contains enough mass to produce emission, typically with $\sigma \lesssim 1$.  This region is characterized by a strong gradient in mass density and entropy, corresponding to a nearly stationary contact discontinuity.  The local Lax-Friedrichs (LLF) Riemann solver employed in {\tt HARM} is known to have poor performance in such flows.  When the gradient is not well resolved, artificial numerical diffusion affects the solution. In our calculations, the $e$-folding length of the entropy is only $\sim 0.15$ cells, i.e., very poorly resolved.   This leads to a largely negative time and $\varphi$-averaged heating rate close to the contact discontinuity, seen in Figure \ref{fig:2Dweight} as the white regions.  Since the entropy equation is only being solved to truncation error while the conservation of energy equation is satisfied to machine precision, it is fine for the \emph{instantaneous} heating rate to be locally negative.  However, it is a concern that the time and spatially integrated heating rate is negative. This still would not be a concern if the negative heating was negligible in magnitude.  However, when integrated over the volume between the black hole event horizon and the inflow equilibrium radius (and limiting the integration domain to $\sigma <1$), the heating totals $\sim - 4.2 \% $ of $ |\dot M c^2|$, roughly the same magnitude as the positive heating ($\sim  4.6  \%$ of $ |\dot M c^2|$).  In Appendix \ref{sec:entwave} we discuss this issue in detail using a simple 1D test of advection of a contact discontinuity.  This test shows explicitly that large, unresolved gradients in entropy lead to negative heating rates such as those seen in Figure \ref{fig:2Dweight}. This is a manifestation of the diffusion of contact discontinuities inherent in finite-volume codes made more extreme by the use of the LLF Riemann solver. These errors do converge to 0 if the contact ``discontinuity'' is actually a smooth but steep transition, but the jet-sheath interface layer is not well-resolved at the current resolution.  It is important to stress that in addition to affecting the heating rate inference, related concerns apply to the thermodynamics of the {\tt HARM} solution as well, which are significantly less accurate in regions of steep (poorly resolved) gradients.

More sophisticated Riemann solvers are known to be significantly less diffusive near contact discontinuities and are particularly well-suited for those with small perpendicular velocity components (as we have here).  It is possible that including one of these solvers (e.g.,  the Harten-Lax-van-Leer-Discontinuities solver) might reduce or remove this negative heating and affect the thermodynamics of the polar outflow.  This in turn might affect the low frequency emission from the simulations.  However, it is not straightforward to implement these more advanced Riemann solvers in a general relativistic framework and thus it has only been done by a handful of groups (e.g., \citealt{Komiss2004,Anton2006,White2016}). We will explore the impact of these more sophisticated Riemann solvers in future work.

\subsection{Low Frequency Radio Slope}
\label{sec:lowfreqrad}

A simple analytic argument can predict the low frequency radio slope for emission from an outflowing plasma \citep{Blandford1979,Falcke1995,Moscijetorig}. Assuming that the magnetic field, mass density, and electron temperature follow power laws in radius of the form:
\begin{align*}
B \propto r^{m_B} \\
\rho \propto r^{m_\rho} \\
\Theta_e \propto r^{m_\Theta},
\end{align*} 
it follows that 
\begin{equation}
F_\nu  \propto \nu^2 \left[\Theta_e A\right](r_{\rm peak}) \propto \left[\Theta_e^5 B^2 A\right](r_{\rm peak}),
\label{Fnu1}
\end{equation}
where $A(r)$ is the cross-sectional area of the outflow at radius $r$ and $r_{\rm peak}$ is the radius at the location of peak emissivity, given by the solution to $\nu \propto [\Theta_e^2 B](r_{\rm peak})$.  Equation \eqref{Fnu1} is simply the black body emission at $r_{\rm peak}(\nu)$ where the optical depth drops to $\sim 1$.  If magnetic flux is conserved, then $B_r A = {\rm const.}$ and $B_\varphi \sqrt{A} = {\rm const.}$, meaning that if $A(r)$ is an increasing function of radius (as it is in conical or parabolic outflows) then $B \approx B_\varphi \propto 1/\sqrt{A}$ at large distances.  Thus we have $F_\nu \propto \Theta_e^5 (r_{\rm peak})$.

Clearly, if the outflow is isothermal, $F_\nu \propto \nu^0$ and the spectrum is flat, matching observations of Sgr A* and other jet sources. On the other hand, if the outflow is adiabatic, i.e. $\Theta_e \propto \rho^{\gamma_e-1}$, then 
\begin{equation}
  F_\nu \propto \nu^{\frac{10 (\gamma_e-1)}{4\gamma_e -3}},
\end{equation}
where we have assumed mass conservation: $\rho A = {\rm const.}$ For $\gamma_e =4/3$, this gives $F_\nu \propto \nu^{10/7}$.  Note that this is independent of the jet shape as long as $A(r)$ increases with radius. 

We find that our simulation generally has an adiabatic outflow (at least in the regions that primarily contribute to the low frequency radio emission) and the low frequency radio slope agrees roughly with $\nu^{10/7}$, under-predicting the observations.  This suggests that our simulations are either missing some important heating mechanism in the outflow or that nonthermal particles (which we do not include) may play a crucial role.  

\section{Comparison to Previous Models of Sgr A*}
\label{sec:comp}

We have shown that our model naturally and self-consistently produces the ``coupled disc-jet'' phenomenological model adopted in previous work to explain observations of Sgr A* \citep{Falcke1995,Yuan2002, Moscijetorig,Mosci2014,CK2015}.   The polar outflow dominates the low frequency radio emission while the accretion disc dominates the higher frequency emission.  Our work thus provides strong theoretical support for some of the assumptions used by previous phenomenological models. However, we also find some important differences which we now highlight for a few representative cases.

\citet{Yuan2002} used a self-consistent analytical model that coupled an ADAF disc to a shocked outflow.  In order to fit the spectrum of Sgr A* they required an electron heating fraction in the disc of $f_e = 10^{-3}$, corresponding to a maximum disc temperature of $\Theta_e \approx 1$ and resulting in $\Theta_e \approx 35$ at the base of the jet.  As a result, the jet dominates the emission at all frequencies except for a narrow region around 230 GHz and a narrow region around $10^{15}$ Hz.  In our model we find that outflow accounts only for the emission below about $10^{11}$ Hz while the disc dominates the emission above $10^{11}$ Hz (see Figure \ref{fig:freqplots}).  The reason for this difference is that although our disc has, on average, $T_e \ll T_g$, there are localized hot-spots of low $\beta$ that contribute significantly to the emission. These hot spots are natural in a turbulent 3D simulation but cannot be captured easily in 1D temperature profiles.  

Disc-jet models in 3D GRMHD simulations have typically assigned a constant relativistic electron temperature, $T_e$, to the ``jet'' and a relatively large constant proton-to-electron temperature ratio, $T_p/T_e$, to the ``disc''.  Two notable examples are \citet{Mosci2014}, who defined the ``jet'' as regions with $-\rho h u_t > 1.02$, where $h$ is the specific relativistic enthalpy, and \citet{CK2015}, who defined the ``jet'' as regions with $\beta<0.2$.  Here ``jet'' and ``disc'' are in quotation marks because both the criterion $\beta <0.2$ and the criterion $-\rho h u_t > 1.02$ occasionally include localized regions close to the horizon in the disc proper (i.e. inflowing material near the midplane).  These small regions in the disc are then assigned relativistic electron temperatures and become a dominant source of higher frequency emission ($\gtrsim 230$ GHz).  Thus,  in contrast to the 1D analytic models, only the low frequency emission ($\lesssim 230 $ GHz) is provided by the outflow while the rest of the emission is provided by the turbulent inflow.  

Despite having similar emitting regions, \citet{Mosci2014} and \cite{CK2015} differ substantially in the amount of NIR emission they predict.  This difference is due to the different choices of electron temperatures for the ``jet'' regions. \citet{Mosci2014} used relatively small electron temperatures ($\Theta_e \sim 10-20$), while \citet{CK2015} used relatively large electron temperatures ($\Theta_e \sim 30-60$ in their SANE models).  The former choice leads to a low level of NIR emission and a dearth of flaring events, contrary to what we find (see Figures \ref{fig:SANEspec} and \ref{fig:Fvt}). The latter choice, on the other hand, is closer to the temperatures of the hot-spots in our model ($\Theta_e \sim 100$; see Figure \ref{fig:freqplots}) and leads to significant NIR emission via frequent flares \citep{Chan2015}, similar to what we find.

 We note, however, that our X-ray emission and X-ray time variability differ substantially from \citet{CK2015} and \citet{Chan2015} due to the different emission processes considered.  Their work neglects Compton scattering and the X-rays in their model are entirely produced by bremsstrahlung at large radii.  On the other hand, our calculations include Compton scattering and focus on the inner portion of the disc where bremsstrahlung is negligible.  We have shown that this Compton component is not only large enough to account for the ``quiescent'' emission but is also a source of weak X-ray flaring. Thus it is crucial to include it for any comparison to X-ray data of Sgr A*.

Our model under-predicts the radio emission of Sgr A* below $\sim 10^{11}$ Hz.  This is, by contrast, well fit by the phenomenological models.  This is because we find the outflow to be roughly adiabatic and not isothermal,  leading to a much steeper spectral slope (see Section \ref{sec:lowfreqrad}).  Electron thermal conduction would seem a natural way for the outflow to be closer to isothermal, but we have shown that it has a negligible effect even with a large thermal conductivity.  This implies that our simulations are either failing to capture enough heating in the outflow or are missing additional physics (e.g., nonthermal particles, heating due to pressure anisotropy, etc.). Our calculation of the heating in the outflow may also be limited by numerical diffusion (see \S \ref{sec:Thermpol} and Appendix \ref{sec:entwave}).  Future work with improved numerical methods and physical models will help to distinguish among these possibilities. 

\begin{figure}
 \begin{center}
\includegraphics[scale = .1]{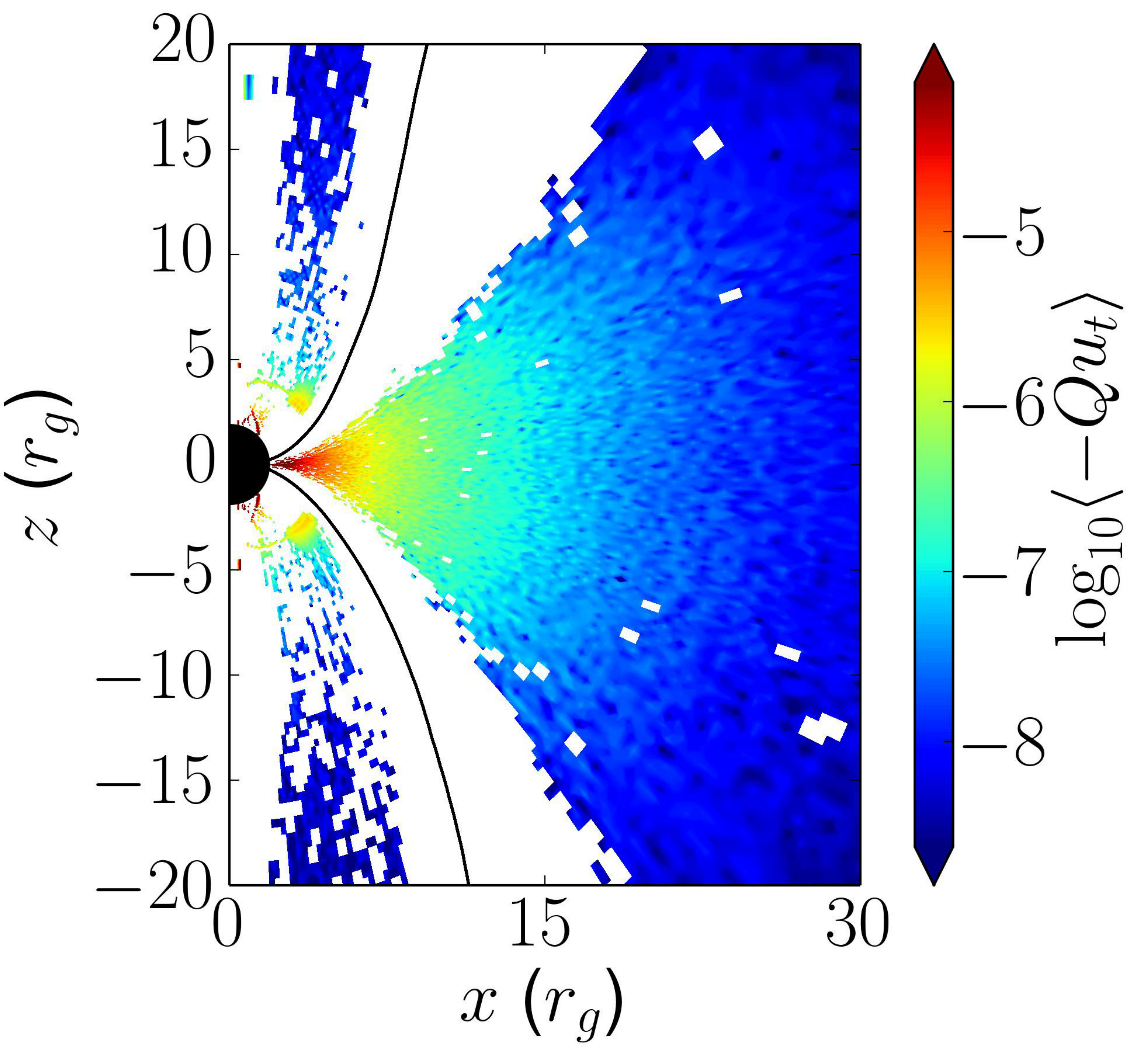}
\end{center}
\caption{Time and $\varphi$-averaged heating rate per unit volume, with the black lines denoting the $b^2/(\rho c^2)$ = 1 contour. The white regions represent negative heating rates, which are unphysical and are a consequence of the LLF Riemann solver's poor treatment of contact discontinuities (the strong density gradient in the polar region; see Figure \ref{fig:SANEFluid}; see \S \ref{sec:Thermpol} and Appendix \ref{sec:entwave}).    Better Riemann solvers should reduce this diffusion and could modify the thermodynamics in the polar region that dominates the low frequency radio emission in our models of Sgr A*.}
\label{fig:2Dweight}
\end{figure}
\section{Conclusions}
\label{sec:conc}

We have presented the first results of applying a self-consistent treatment of electron thermodynamics in GRMHD simulations to observations of Sgr A*.  Our goal is not to fit the observations in detail but to qualitatively determine the basic predictions of our model in which the electron entropy is self-consistently evolved including electron heating and anisotropic electron thermal conduction. In our calculation, the electron heating fraction, $f_e$, is not a free parameter adjusted to match observations but is fixed by a first principles calculation (\S \ref{sec:fluid}).  We have only modeled the emission by thermal electrons, deferring models of non-thermal particle acceleration and emission to future work

Despite the lack of any free parameters tuned to match observations, we find encouraging agreement between some properties of our predicted spectra and variability and observations of Sgr A*. This agreement includes the spectral slope near $230$ GHz and the approximate magnitude of the time-averaged NIR and X-ray emission (Figure \ref{fig:SANEspec}), as well as many of the qualitative features of the time variability (Figure \ref{fig:Fvt}).  The images produced by our model at 230 GHz and 2 $\mu$m (Figure \ref{fig:SANEimage}) display the characteristic ``Einstein ring''  at all inclination angles. This is encouraging for upcoming horizon-scale observations at these frequencies.

Our model reproduces the disc-jet structure of the accretion flow appealed to in the literature to explain the emission from Sgr A* \citep{Yuan2002,Mosci2014,CK2015}. The polar outflow dominates the low frequency radio emission and the turbulent inflow dominates the emission at frequencies above $\sim 230$ GHz.   This can be seen directly using the properties of the emitting regions as a function of frequency in Figure \ref{fig:freqplots}, and the 30 GHz, 230 GHz, and 2 $\mu$m images in Figure \ref{fig:SANEimage}.  Our results thus provide physical justification for many of the phenomenological prescriptions used in the past to model Sgr A* and can help motivate more sophisticated variants of these models in the future.

We find that anisotropic electron thermal conduction along magnetic field lines has little effect on our model spectra and images (see Figure \ref{fig:SANEspec} for the SEDs; images including conduction are not shown here because they are nearly identical to those without conduction in Figure \ref{fig:SANEimage}).  This is because the field lines in our simulation are primarily toroidal while the temperature gradients are primarily poloidal. \citet{Foucart2016} came to a similar conclusion for the effect of ion conduction on the fluid dynamics when including it in the total fluid stress-energy tensor. The small effect of electron conduction in our model suggests that future modelling of electron thermodynamics can probably neglect electron conduction, greatly simplifying the numerics and reducing the computational expense by a factor of $\sim 4$. This needs, however, to be confirmed for Magnetically Arrested Discs (MADs, \citealt{2003PASJ...55L..69N,Sasha2011}) or any other disc which differs significantly from the structure in our model.

The variability in our simulations is qualitatively similar to that seen in Sgr A* (Figure \ref{fig:Fvt}): 1) we find frequent X-ray and NIR flaring events 2) those flares are correlated 3) mm emission displays less variability 4) all X-ray flares have NIR counterparts, but we also find evidence for NIR flares without X-ray counterparts. The X-ray flares, however, are much weaker in magnitude than the strong flaring events observed in Sgr A*. This could be a result of our relatively small sample size in time ($\sim$ 1 day) or it could imply the need for nonthermal particles (e.g. \citealt{Ozel2000,Yuan2003,Ball2016}).   Given the wealth of time-dependent observational data on Sgr A* and the uncertainties that remain in its underlying physics, it is of great interest to study the statistical properties of the model's time variability in more detail. This will be a focus of future work.

Finally, like all numerical calculations, our model is not without its potential failure modes.  In our disc models, numerical diffusion of poorly resolved entropy gradients affects the thermodynamics of both the electrons and the total fluid in the polar regions.  This may affect the low frequency ($\lesssim 10^{11}$ Hz)  emission in our model.  On the other hand, all emission above $\sim 10^{11}$ Hz is unaffected by this issue.  Moreover, we find that doubling the resolution of our simulation does not affect our predicted spectra (see Figure \ref{fig:spec_convergence}), but somewhat improves the agreement of the low-frequency slope with the observations.

\section*{Acknowledgments}
We thank B. Ryan, J. Neilsen, A. Sadowski, and R. Narayan for useful discussions, as well as all the members of the horizon collaboration, \href{http://horizon.astro.illinois.edu}{http://horizon.astro.illinois.edu}, for their advice and encouragement.   We also thank the referee, F. Yuan, for useful comments on the manuscript. This work was supported by NSF grant AST 13-33612, and a Romano Professorial Scholar appointment to CFG.   SMR is supported in part by the NASA Earth and Space Science Fellowship.  EQ is supported in part by a Simons Investigator Award from the Simons Foundation and the David and Lucile Packard Foundation.   Support for AT was provided by NASA through Einstein Postdoctoral Fellowship grant number PF3-140131 awarded by the Chandra X-ray Center, which is operated by the Smithsonian Astrophysical Observatory for NASA under contract NAS8-03060, the TAC fellowship, and by NSF through an XSEDE computational time allocation TG-AST100040 on TACC Stampede. This work was made possible by computing time granted by UCB on the Savio cluster.

\bibliographystyle{mn2efix}
\bibliography{sgra}

\begin{thebibliography}{73}
\expandafter\ifx\csname natexlab\endcsname\relax\def\natexlab#1{#1}\fi

\bibitem[{{An} {et~al}\mbox{.}(2005){An}, {Goss}, {Zhao}, {Hong}, {Roy}, {Rao},
  \& {Shen}}]{An2005}
{An} T., {Goss} W.~M., {Zhao} J.-H., {Hong} X.~Y., {Roy} S., {Rao} A.~P.,
  {Shen} Z.-Q., 2005, ApJl, 634, L49

\bibitem[{{Ant{\'o}n} {et~al}\mbox{.}(2006){Ant{\'o}n}, {Zanotti}, {Miralles},
  {Mart{\'{\i}}}, {Ib{\'a}{\~n}ez}, {Font}, \& {Pons}}]{Anton2006}
{Ant{\'o}n} L., {Zanotti} O., {Miralles} J.~A., {Mart{\'{\i}}} J.~M.,
  {Ib{\'a}{\~n}ez} J.~M., {Font} J.~A., {Pons} J.~A., 2006, ApJ, 637, 296

\bibitem[{{Baganoff} {et~al}\mbox{.}(2003){Baganoff}, {Maeda}, {Morris},
  {Bautz}, {Brandt}, {Cui}, {Doty}, {Feigelson}, {Garmire}, {Pravdo}, {Ricker},
  \& {Townsley}}]{Baganoff2003}
{Baganoff} F.~K. {et~al.}, 2003, ApJ, 591, 891

\bibitem[{{Balbus} \& {Hawley}(1991)}]{1991ApJ...376..214B}
{Balbus} S.~A., {Hawley} J.~F., 1991, ApJ, 376, 214

\bibitem[{{Ball} {et~al}\mbox{.}(2016){Ball}, {{\"O}zel}, {Psaltis}, \&
  {Chan}}]{Ball2016}
{Ball} D., {{\"O}zel} F., {Psaltis} D., {Chan} C.-k., 2016, \apj, 826, 77

\bibitem[{{Blandford} \& {K{\"o}nigl}(1979)}]{Blandford1979}
{Blandford} R.~D., {K{\"o}nigl} A., 1979, ApJ, 232, 34

\bibitem[{{Blandford} \& {Payne}(1982)}]{BPJet}
{Blandford} R.~D., {Payne} D.~G., 1982, MNRAS, 199, 883

\bibitem[{{Blandford} \& {Znajek}(1977)}]{BZJet}
{Blandford} R.~D., {Znajek} R.~L., 1977, MNRAS, 179, 433

\bibitem[{{Bower} {et~al}\mbox{.}(2015){Bower}, {Markoff}, {Dexter}, {Gurwell},
  {Moran}, {Brunthaler}, {Falcke}, {Fragile}, {Maitra}, {Marrone}, {Peck},
  {Rushton}, \& {Wright}}]{Bower2015}
{Bower} G.~C. {et~al.}, 2015, ApJ, 802, 69

\bibitem[{{Chan} {et~al}\mbox{.}(2015{\natexlab{a}}){Chan}, {Psaltis},
  {{\"O}zel}, {Medeiros}, {Marrone}, {Sa{\c d}owski}, \& {Narayan}}]{Chan2015}
{Chan} C.-K., {Psaltis} D., {{\"O}zel} F., {Medeiros} L., {Marrone} D., {Sa{\c
  d}owski} A., {Narayan} R., 2015{\natexlab{a}}, ApJ, 812, 103

\bibitem[{{Chan} {et~al}\mbox{.}(2015{\natexlab{b}}){Chan}, {Psaltis},
  {{\"O}zel}, {Narayan}, \& {Sa{\c d}owski}}]{CK2015}
{Chan} C.-K., {Psaltis} D., {{\"O}zel} F., {Narayan} R., {Sa{\c d}owski} A.,
  2015{\natexlab{b}}, ApJ, 799, 1

\bibitem[{{Chandra} {et~al}\mbox{.}(2015){Chandra}, {Gammie}, {Foucart}, \&
  {Quataert}}]{ManiModel}
{Chandra} M., {Gammie} C.~F., {Foucart} F., {Quataert} E., 2015, ApJ, 810, 162

\bibitem[{{Cotera} {et~al}\mbox{.}(1999){Cotera}, {Morris}, {Ghez}, {Becklin},
  {Tanner}, {Werner}, \& {Stolovy}}]{Cotera1999}
{Cotera} A., {Morris} M., {Ghez} A.~M., {Becklin} E.~E., {Tanner} A.~M.,
  {Werner} M.~W., {Stolovy} S.~R., 1999, in Astronomical Society of the Pacific
  Conference Series, Vol. 186, The Central Parsecs of the Galaxy, {Falcke} H.,
  {Cotera} A., {Duschl} W.~J., {Melia} F., {Rieke} M.~J., eds., p. 240

\bibitem[{{De Villiers} \& {Hawley}(2003)}]{DeVilliers2003}
{De Villiers} J.-P., {Hawley} J.~F., 2003, ApJ, 589, 458

\bibitem[{{Do} {et~al}\mbox{.}(2009){Do}, {Ghez}, {Morris}, {Yelda}, {Meyer},
  {Lu}, {Hornstein}, \& {Matthews}}]{Do2009}
{Do} T., {Ghez} A.~M., {Morris} M.~R., {Yelda} S., {Meyer} L., {Lu} J.~R.,
  {Hornstein} S.~D., {Matthews} K., 2009, ApJ, 691, 1021

\bibitem[{{Doeleman} {et~al}\mbox{.}(2009){Doeleman}, {Agol}, {Backer},
  {Baganoff}, {Bower}, {Broderick}, {Fabian}, {Fish}, {Gammie}, {Ho}, {Honman},
  {Krichbaum}, {Loeb}, {Marrone}, {Reid}, {Rogers}, {Shapiro}, {Strittmatter},
  {Tilanus}, {Weintroub}, {Whitney}, {Wright}, \& {Ziurys}}]{Doeleman2009}
{Doeleman} S. {et~al.}, 2009, in Astronomy, Vol. 2010, astro2010: The Astronomy
  and Astrophysics Decadal Survey, p.~68

\bibitem[{{Doeleman} {et~al}\mbox{.}(2008){Doeleman}, {Weintroub}, {Rogers},
  {Plambeck}, {Freund}, {Tilanus}, {Friberg}, {Ziurys}, {Moran}, {Corey},
  {Young}, {Smythe}, {Titus}, {Marrone}, {Cappallo}, {Bock}, {Bower},
  {Chamberlin}, {Davis}, {Krichbaum}, {Lamb}, {Maness}, {Niell}, {Roy},
  {Strittmatter}, {Werthimer}, {Whitney}, \& {Woody}}]{Doeleman2008}
{Doeleman} S.~S. {et~al.}, 2008, Nature, 455, 78

\bibitem[{{Dolence} {et~al}\mbox{.}(2009){Dolence}, {Gammie},
  {Mo{\'s}cibrodzka}, \& {Leung}}]{Dolence2009}
{Dolence} J.~C., {Gammie} C.~F., {Mo{\'s}cibrodzka} M., {Leung} P.~K., 2009,
  ApJs, 184, 387

\bibitem[{{Eckart} {et~al}\mbox{.}(2012){Eckart}, {Garc{\'{\i}}a-Mar{\'{\i}}n},
  {Vogel}, {Teuben}, {Morris}, {Baganoff}, {Dexter}, {Sch{\"o}del}, {Witzel},
  {Valencia-S.}, {Karas}, {Kunneriath}, {Straubmeier}, {Moser}, {Sabha},
  {Buchholz}, {Zamaninasab}, {Mu{\v z}i{\'c}}, {Moultaka}, \&
  {Zensus}}]{Eckart2012}
{Eckart} A. {et~al.}, 2012, A \& A, 537, A52

\bibitem[{{Falcke} \& {Biermann}(1995)}]{Falcke1995}
{Falcke} H., {Biermann} P.~L., 1995, A\&A, 293, 665

\bibitem[{{Falcke} {et~al}\mbox{.}(1998){Falcke}, {Goss}, {Matsuo}, {Teuben},
  {Zhao}, \& {Zylka}}]{Falcke1998}
{Falcke} H., {Goss} W.~M., {Matsuo} H., {Teuben} P., {Zhao} J.-H., {Zylka} R.,
  1998, ApJ, 499, 731

\bibitem[{{Fishbone} \& {Moncrief}(1976)}]{Fishbone1976}
{Fishbone} L.~G., {Moncrief} V., 1976, ApJ, 207, 962

\bibitem[{{Foucart} {et~al}\mbox{.}(2016){Foucart}, {Chandra}, {Gammie}, \&
  {Quataert}}]{Foucart2016}
{Foucart} F., {Chandra} M., {Gammie} C.~F., {Quataert} E., 2016, MNRAS, 456,
  1332

\bibitem[{{Gammie}, {McKinney} \& {T{\'o}th}(2003){Gammie}, {McKinney}, \&
  {T{\'o}th}}]{Gammie2003}
{Gammie} C.~F., {McKinney} J.~C., {T{\'o}th} G., 2003, ApJ, 589, 444

\bibitem[{{Genzel} \& {Eckart}(1999)}]{Genzel1999}
{Genzel} R., {Eckart} A., 1999, in Astronomical Society of the Pacific
  Conference Series, Vol. 186, The Central Parsecs of the Galaxy, {Falcke} H.,
  {Cotera} A., {Duschl} W.~J., {Melia} F., {Rieke} M.~J., eds., p.~3

\bibitem[{{Genzel} {et~al}\mbox{.}(2003){Genzel}, {Sch{\"o}del}, {Ott},
  {Eckart}, {Alexander}, {Lacombe}, {Rouan}, \& {Aschenbach}}]{Genzel2003}
{Genzel} R., {Sch{\"o}del} R., {Ott} T., {Eckart} A., {Alexander} T., {Lacombe}
  F., {Rouan} D., {Aschenbach} B., 2003, Nature, 425, 934

\bibitem[{{Gillessen} {et~al}\mbox{.}(2010){Gillessen}, {Eisenhauer}, {Perrin},
  {Brandner}, {Straubmeier}, {Perraut}, {Amorim}, {Sch{\"o}ller},
  {Araujo-Hauck}, {Bartko}, {Baumeister}, {Berger}, {Carvas}, {Cassaing},
  {Chapron}, {Choquet}, {Clenet}, {Collin}, {Eckart}, {Fedou}, {Fischer},
  {Gendron}, {Genzel}, {Gitton}, {Gonte}, {Gr{\"a}ter}, {Haguenauer}, {Haug},
  {Haubois}, {Henning}, {Hippler}, {Hofmann}, {Jocou}, {Kellner}, {Kervella},
  {Klein}, {Kudryavtseva}, {Lacour}, {Lapeyrere}, {Laun}, {Lena}, {Lenzen},
  {Lima}, {Moratschke}, {Moch}, {Moulin}, {Naranjo}, {Neumann}, {Nolot},
  {Paumard}, {Pfuhl}, {Rabien}, {Ramos}, {Rees}, {Rohloff}, {Rouan}, {Rousset},
  {Sevin}, {Thiel}, {Wagner}, {Wiest}, {Yazici}, \& {Ziegler}}]{Gillessen2010}
{Gillessen} S. {et~al.}, 2010, in Society of Photo-Optical Instrumentation
  Engineers (SPIE) Conference Series, Vol. 7734, Society of Photo-Optical
  Instrumentation Engineers (SPIE) Conference Series, p.~0

\bibitem[{{Hornstein} {et~al}\mbox{.}(2007){Hornstein}, {Matthews}, {Ghez},
  {Lu}, {Morris}, {Becklin}, {Rafelski}, \& {Baganoff}}]{Hornstein2007}
{Hornstein} S.~D., {Matthews} K., {Ghez} A.~M., {Lu} J.~R., {Morris} M.,
  {Becklin} E.~E., {Rafelski} M., {Baganoff} F.~K., 2007, ApJ, 667, 900

\bibitem[{{Howes}(2010)}]{Howes2010}
{Howes} G.~G., 2010, MNRAS, 409, L104

\bibitem[{{Howes}(2011)}]{Howes2011}
{Howes} G.~G., 2011, ApJ, 738, 40

\bibitem[{{Ichimaru}(1977)}]{Ichimaru1977}
{Ichimaru} S., 1977, ApJ, 214, 840

\bibitem[{{Komissarov}(1999)}]{Komiss1999}
{Komissarov} S.~S., 1999, MNRAS, 303, 343

\bibitem[{{Komissarov}(2004)}]{Komiss2004}
{Komissarov} S.~S., 2004, MNRAS, 350, 1431

\bibitem[{{Liu} {et~al}\mbox{.}(2016{\natexlab{a}}){Liu}, {Wright}, {Zhao},
  {Brinkerink}, {Ho}, {Mills}, {Mart{\'{\i}}n}, {Falcke}, {Matsushita}, \&
  {Mart{\'{\i}}-Vidal}}]{Liu2016}
{Liu} H.~B. {et~al.}, 2016{\natexlab{a}}, A \& A, 593, A107

\bibitem[{{Liu} {et~al}\mbox{.}(2016{\natexlab{b}}){Liu}, {Wright}, {Zhao},
  {Mills}, {Requena-Torres}, {Matsushita}, {Mart{\'{\i}}n}, {Ott}, {Morris},
  {Longmore}, {Brinkerink}, \& {Falcke}}]{Liu2016492GHz}
{Liu} H.~B. {et~al.}, 2016{\natexlab{b}}, A \& A, 593, A44

\bibitem[{{Lynden-Bell}(2003)}]{LyndenBell2003}
{Lynden-Bell} D., 2003, MNRAS, 341, 1360

\bibitem[{{Marrone} {et~al}\mbox{.}(2007){Marrone}, {Moran}, {Zhao}, \&
  {Rao}}]{Marrone2007}
{Marrone} D.~P., {Moran} J.~M., {Zhao} J.-H., {Rao} R., 2007, \apjl, 654, L57

\bibitem[{{McKinney} \& {Gammie}(2004)}]{McKinney2004}
{McKinney} J.~C., {Gammie} C.~F., 2004, ApJ, 611, 977

\bibitem[{{McKinney}, {Tchekhovskoy} \& {Blandford}(2012){McKinney},
  {Tchekhovskoy}, \& {Blandford}}]{2012MNRAS.423.3083M}
{McKinney} J.~C., {Tchekhovskoy} A., {Blandford} R.~D., 2012, \mnras, 423, 3083

\bibitem[{{Mo{\'s}cibrodzka} \& {Falcke}(2013)}]{Moscijetorig}
{Mo{\'s}cibrodzka} M., {Falcke} H., 2013, A\&A, 559, L3

\bibitem[{{Mo{\'s}cibrodzka} {et~al}\mbox{.}(2014){Mo{\'s}cibrodzka}, {Falcke},
  {Shiokawa}, \& {Gammie}}]{Mosci2014}
{Mo{\'s}cibrodzka} M., {Falcke} H., {Shiokawa} H., {Gammie} C.~F., 2014, A\&A,
  570, A7

\bibitem[{{Mo{\'s}cibrodzka} {et~al}\mbox{.}(2009){Mo{\'s}cibrodzka}, {Gammie},
  {Dolence}, {Shiokawa}, \& {Leung}}]{Mosci2009}
{Mo{\'s}cibrodzka} M., {Gammie} C.~F., {Dolence} J.~C., {Shiokawa} H., {Leung}
  P.~K., 2009, ApJ, 706, 497

\bibitem[{{Narayan}, {Igumenshchev} \& {Abramowicz}(2003){Narayan},
  {Igumenshchev}, \& {Abramowicz}}]{2003PASJ...55L..69N}
{Narayan} R., {Igumenshchev} I.~V., {Abramowicz} M.~A., 2003, \pasj, 55, L69

\bibitem[{{Narayan} {et~al}\mbox{.}(2012){Narayan}, {S{\"A} dowski}, {Penna},
  \& {Kulkarni}}]{2012MNRAS.426.3241N}
{Narayan} R., {S{\"A} dowski} A., {Penna} R.~F., {Kulkarni} A.~K., 2012,
  \mnras, 426, 3241

\bibitem[{{Narayan} \& {Yi}(1994)}]{1994ApJ...428L..13N}
{Narayan} R., {Yi} I., 1994, \apjl, 428, L13

\bibitem[{{Neilsen} {et~al}\mbox{.}(2013){Neilsen}, {Nowak}, {Gammie},
  {Dexter}, {Markoff}, {Haggard}, {Nayakshin}, {Wang}, {Grosso}, {Porquet},
  {Tomsick}, {Degenaar}, {Fragile}, {Houck}, {Wijnands}, {Miller}, \&
  {Baganoff}}]{Neilsen2013}
{Neilsen} J. {et~al.}, 2013, ApJ, 774, 42

\bibitem[{{Noble} {et~al}\mbox{.}(2006){Noble}, {Gammie}, {McKinney}, \& {Del
  Zanna}}]{2006ApJ...641..626N}
{Noble} S.~C., {Gammie} C.~F., {McKinney} J.~C., {Del Zanna} L., 2006, ApJ,
  641, 626

\bibitem[{{Noble} {et~al}\mbox{.}(2007){Noble}, {Leung}, {Gammie}, \&
  {Book}}]{Noble2007}
{Noble} S.~C., {Leung} P.~K., {Gammie} C.~F., {Book} L.~G., 2007, Classical and
  Quantum Gravity, 24, 259

\bibitem[{{Novikov} \& {Thorne}(1973)}]{Novikov1973}
{Novikov} I.~D., {Thorne} K.~S., 1973, in Black Holes (Les Astres Occlus),
  {Dewitt} C., {Dewitt} B.~S., eds., pp. 343--450

\bibitem[{{Numata} \& {Loureiro}(2015)}]{Numata2014}
{Numata} R., {Loureiro} N.~F., 2015, Journal of Plasma Physics, 81, 023001

\bibitem[{{\"O}zel, Psaltis \& Narayan(2000){\"O}zel, Psaltis, \&
  Narayan}]{Ozel2000}
{\"O}zel F., Psaltis D., Narayan R., 2000, ApJ, 541, 234

\bibitem[{{Quataert}(2001)}]{2001ASPC..224...71Q}
{Quataert} E., 2001, in Astronomical Society of the Pacific Conference Series,
  Vol. 224, Probing the Physics of Active Galactic Nuclei, {Peterson} B.~M.,
  {Pogge} R.~W., {Polidan} R.~S., eds., p.~71

\bibitem[{{Quataert} \& {Gruzinov}(1999)}]{Quataert1999}
{Quataert} E., {Gruzinov} A., 1999, ApJ, 520, 248

\bibitem[{{Rees} {et~al}\mbox{.}(1982){Rees}, {Begelman}, {Blandford}, \&
  {Phinney}}]{Rees1982}
{Rees} M.~J., {Begelman} M.~C., {Blandford} R.~D., {Phinney} E.~S., 1982,
  Nature, 295, 17

\bibitem[{{Ressler} {et~al}\mbox{.}(2015){Ressler}, {Tchekhovskoy}, {Quataert},
  {Chandra}, \& {Gammie}}]{Ressler2015}
{Ressler} S.~M., {Tchekhovskoy} A., {Quataert} E., {Chandra} M., {Gammie}
  C.~F., 2015, MNRAS, 454, 1848

\bibitem[{{Sadowski} {et~al}\mbox{.}(2016){Sadowski}, {Wielgus}, {Narayan},
  {Abarca}, {McKinney}, \& {Chael}}]{Sadowski2016}
{Sadowski} A., {Wielgus} M., {Narayan} R., {Abarca} D., {McKinney} J.~C.,
  {Chael} A., 2016, ArXiv e-prints

\bibitem[{{Sch{\"o}del} {et~al}\mbox{.}(2007){Sch{\"o}del}, {Eckart}, {Mu{\v
  z}i{\'c}}, {Meyer}, {Viehmann}, \& {Bower}}]{Schodel2007}
{Sch{\"o}del} R., {Eckart} A., {Mu{\v z}i{\'c}} K., {Meyer} L., {Viehmann} T.,
  {Bower} G.~C., 2007, A\&A, 462, L1

\bibitem[{{Sch{\"o}del} {et~al}\mbox{.}(2011){Sch{\"o}del}, {Morris}, {Muzic},
  {Alberdi}, {Meyer}, {Eckart}, \& {Gezari}}]{Schodel2011}
{Sch{\"o}del} R., {Morris} M.~R., {Muzic} K., {Alberdi} A., {Meyer} L.,
  {Eckart} A., {Gezari} D.~Y., 2011, A\&A, 532, A83

\bibitem[{{Shcherbakov} \& {Baganoff}(2010)}]{Shcherbakov2010}
{Shcherbakov} R.~V., {Baganoff} F.~K., 2010, ApJ, 716, 504

\bibitem[{{Shcherbakov}, {Penna} \& {McKinney}(2012){Shcherbakov}, {Penna}, \&
  {McKinney}}]{Shcherbakov2012}
{Shcherbakov} R.~V., {Penna} R.~F., {McKinney} J.~C., 2012, ApJ, 755, 133

\bibitem[{{Tchekhovskoy}, {McKinney} \& {Narayan}(2007){Tchekhovskoy},
  {McKinney}, \& {Narayan}}]{Sasha2007}
{Tchekhovskoy} A., {McKinney} J.~C., {Narayan} R., 2007, MNRAS, 379, 469

\bibitem[{{Tchekhovskoy}, {McKinney} \& {Narayan}(2009){Tchekhovskoy},
  {McKinney}, \& {Narayan}}]{2009ApJ...699.1789T}
{Tchekhovskoy} A., {McKinney} J.~C., {Narayan} R., 2009, \apj, 699, 1789

\bibitem[{{Tchekhovskoy}, {Narayan} \& {McKinney}(2010){Tchekhovskoy},
  {Narayan}, \& {McKinney}}]{2010ApJ...711...50T}
{Tchekhovskoy} A., {Narayan} R., {McKinney} J.~C., 2010, ApJ, 711, 50

\bibitem[{{Tchekhovskoy}, {Narayan} \& {McKinney}(2011){Tchekhovskoy},
  {Narayan}, \& {McKinney}}]{Sasha2011}
{Tchekhovskoy} A., {Narayan} R., {McKinney} J.~C., 2011, MNRAS, 418, L79

\bibitem[{{Telesco}, {Davidson} \& {Werner}(1996){Telesco}, {Davidson}, \&
  {Werner}}]{Telesco1996}
{Telesco} C.~M., {Davidson} J.~A., {Werner} M.~W., 1996, ApJ, 456, 541

\bibitem[{{Toro}(2009)}]{Toro2009}
{Toro} E.~F., 2009, {Riemann Solvers and Numerical Methods for Fluid Dynamics}.
  Springer

\bibitem[{{Trap} {et~al}\mbox{.}(2011){Trap}, {Goldwurm}, {Dodds-Eden},
  {Weiss}, {Terrier}, {Ponti}, {Gillessen}, {Genzel}, {Ferrando},
  {B{\'e}langer}, {Cl{\'e}net}, {Rouan}, {Predehl}, {Capelli}, {Melia}, \&
  {Yusef-Zadeh}}]{Trap2011}
{Trap} G. {et~al.}, 2011, A \& A, 528, A140

\bibitem[{{White}, {Stone} \& {Gammie}(2016){White}, {Stone}, \&
  {Gammie}}]{White2016}
{White} C.~J., {Stone} J.~M., {Gammie} C.~F., 2016, ApJs, 225, 22

\bibitem[{{Yuan}, {Bu} \& {Wu}(2012){Yuan}, {Bu}, \& {Wu}}]{Yuan2012}
{Yuan} F., {Bu} D., {Wu} M., 2012, ApJ, 761, 130

\bibitem[{{Yuan} {et~al}\mbox{.}(2015){Yuan}, {Gan}, {Narayan}, {Sadowski},
  {Bu}, \& {Bai}}]{Yuan2015}
{Yuan} F., {Gan} Z., {Narayan} R., {Sadowski} A., {Bu} D., {Bai} X.-N., 2015,
  ApJ, 804, 101

\bibitem[{{Yuan}, {Markoff} \& {Falcke}(2002){Yuan}, {Markoff}, \&
  {Falcke}}]{Yuan2002}
{Yuan} F., {Markoff} S., {Falcke} H., 2002, A\&A, 383, 854

\bibitem[{{Yuan} \& {Narayan}(2014)}]{2014ARA&A..52..529Y}
{Yuan} F., {Narayan} R., 2014, \araa, 52, 529

\bibitem[{{Yuan}, {Quataert} \& {Narayan}(2003){Yuan}, {Quataert}, \&
  {Narayan}}]{Yuan2003}
{Yuan} F., {Quataert} E., {Narayan} R., 2003, ApJ, 598, 301

\end{thebibliography}

\appendix

\section{Observational Data}
\label{sec:obs}
The radio and millimetre data we use are taken from the mean spectra of Sagittarius A* as calculated by  \citet{Falcke1998}, combining the Very Large Array (VLA), the Institut de Radioastronomie Millimetrique (IRAM), the Nobeyama 45 m, and the Berkeley-Illinois-Maryland Array (BIMA) observational results over the frequency interval 1.46 - 235.6 GHz, \citet{An2005} combining VLA and Giant Metrewave Radio Telescope (GMRT) observational results over the frequency interval 0.33 - 42.9 GHz, \citet{Doeleman2008} using the Event Horizon Telescope (EHT) at 230 GHz, \citet{Bower2015}, combining VLA, the Atacama Large Millimetre/submillimetre Array (ALMA), and the Submillimetre Array (SMA) observational results over the frequency interval 1.6 - 352.6 GHz, and finally, \citet{Liu2016} and \citet{Liu2016492GHz} using ALMA over the frequency interval 92.996 - 708.860 GHz and at 492 GHz, respectively.  Infrared upper limits are taken from \citet{Telesco1996} at 30 $\mu$m, \citet{Cotera1999} at 24.5 $\mu$m and 8.81 $\mu$m, \citet{Genzel1999} at 2.2 $\mu$m, and \citet{Schodel2007} at 8.6 $\mu$m, while infrared data points for both the quiescent and flare states are taken from \citet{Genzel2003} and \citet{Schodel2011}.  Finally, we use the range of 2-10 keV X-ray flare luminosities seen in Sgr A* during the year 2012 \citep{Neilsen2013} and the quiescent level of 2.4 $
\times 10^{33}$ erg s$^{-1}$ \citep{Baganoff2003}. Note that the inner accretion flow is believed to contribute only $\sim 10 \%$ of this quiescent flux \citep{Neilsen2013}, but we use the total quiescent emission to define the luminosity that ``flares'' must exceed in our model.

\section{Initial conditions, Coordinate System and Floors }
\label{App:coords}

\begin{figure}
    \includegraphics[width=\columnwidth]{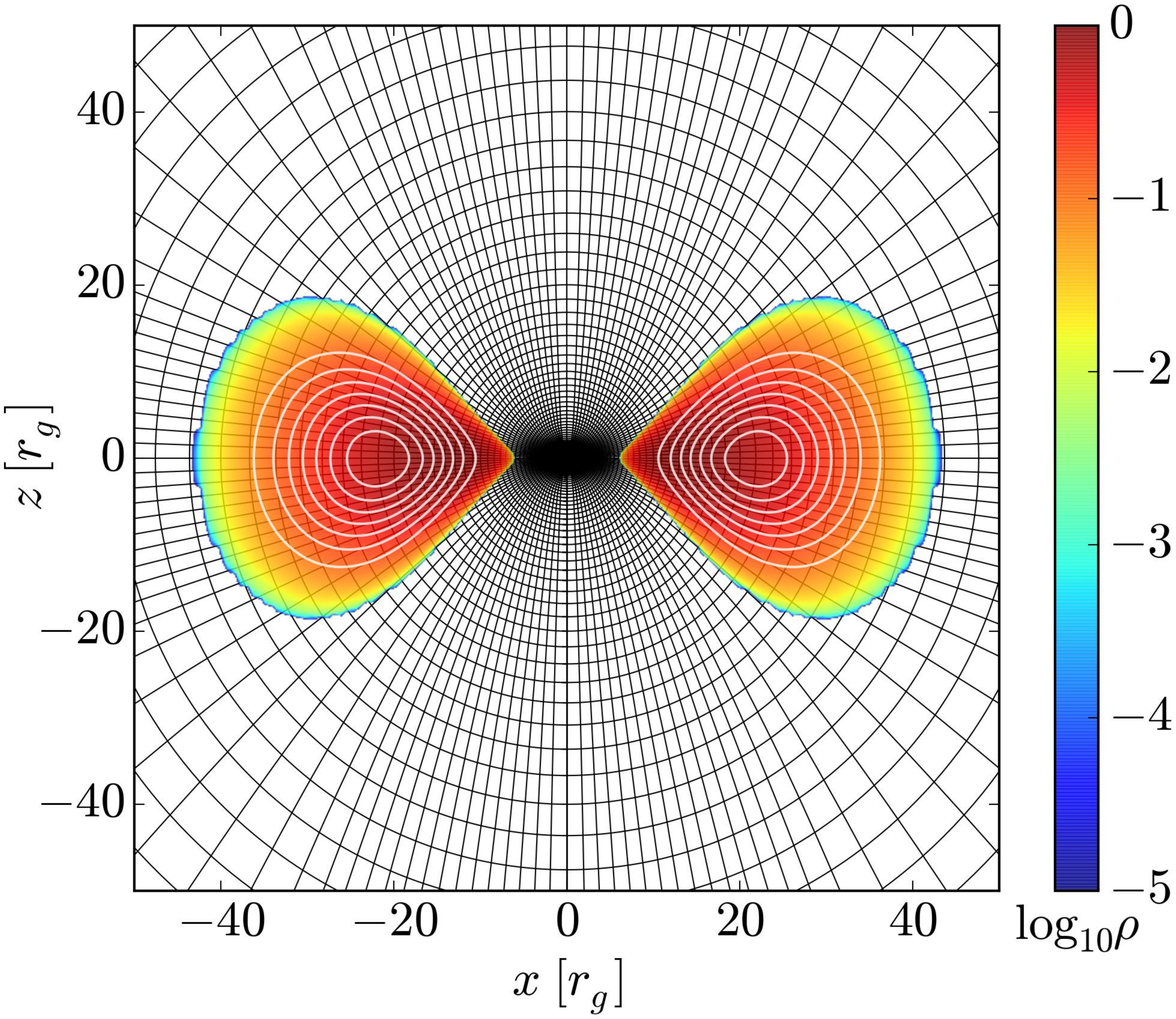} 
\caption{Our simulations use a grid adapted for the disc-jet problem. The grid lines, shown with black lines (every 4th cell interface is shown for our fiducial model at the resolution of $320\times256\times64$), concentrate resolution both toward the equatorial plane to resolve the disc turbulence and toward the polar regions to resolve the twin polar jets. Colours show the initial density distribution on a logarithmic scale (red shows high, blue low values, please see the colour bar). White lines show initial poloidal magnetic field lines.}%
\label{fig:ic_grid}
\end{figure}

In this appendix we describe our specific choice of initial conditions, coordinates, and numerical floors.

\subsection{Initial Conditions}
One approach for initial conditions is to overlay the torus with a weak magnetic field with field lines along contours of density,
$A_\varphi \propto \rho-\rho_0$ \citep{Gammie2003}. This results in a single magnetic field loop in the $(r,\theta)$ plane. The centre of the loop is at the density maximum, and the loop is fully contained within the torus. However, we found that by about $t\sim 10^4r_g/c$ such a loop gets nearly fully consumed by the black hole.  To avoid this, we opt for a larger loop that survives for a longer time. We choose magnetic vector potential $A_\varphi \propto r^4\rho^2$. Figure~\ref{fig:ic_grid} shows that the radial pre-factor shifts the loop to larger radii and increases its size: the centre of the loop shifts from $r=13r_g$ to $\sim22r_g$. This leads to a stronger initial magnetic field at larger radii and makes it easier to resolve the magnetorotational instability (MRI, \citealt{1991ApJ...376..214B}) throughout the torus. We normalize the magnetic field strength such that the ratio of maximum gas to maximum magnetic pressure equals $100$.

\subsection{Coordinate System}

We use a variant of 3D modified Kerr-Schild (MKS) coordinates \citep{Gammie2003}. {\tt HARM} discretizes the equations of motion on a uniform grid in the internal code coordinates, $t, x^{(1)}, x^{(2)}, x^{(3)}$. Figure \ref{fig:ic_grid} displays this grid in Boyer-Lindquist coordinates with the initial conditions overplotted on top of it.  

{\it Radial grid.} The original MKS coordinates used an exponential mapping of the internal radial coordinate $x^{(1)}$ into radius $r$: $r= \exp[x^{(1)}]$.  In this work, we instead use a ``hyper-exponential'' mapping \citep{Sasha2011}, as described in previous work \citep{2009ApJ...699.1789T,Sasha2011,Ressler2015}. Outside a break radius, $r_{\rm br}\equiv\exp(x_{\rm br})$, the radial grid becomes highly unresolved and spans very a large distance in just a few cells: $r = \exp[x^{(1)}+4(x^{(1)}-x_{\rm br})^4]$.  Inside $r_{\rm br}$, this radial mapping is equivalent to the MKS coordinates.  The advantage of this grid is that it prevents unphysical reflections off the outer radial grid boundary by moving the boundary out of causal contact with the disc, outflows, and the jets. Here we take $r_{\rm br} = 400 M$, which guarantees that the regions of interest $r\lesssim r_{\rm br}$ are well-resolved.

{\it Angular grid.} The original MKS coordinates used non-collimating grid lines that followed lines of $\theta = {\rm const}$. Since we are interested in the physics of the jet sheath, which collimates into small opening angles, we adopt a grid that focuses substantial resolution into the polar regions of interest. The grid consists of disc and jet angular patches that are smoothly stitched together using a transition function, 
\begin{equation}
\Theta_s(x,x_a,x_b,y_a,y_b) = y_a+(y_b-y_a)\, \tilde\Theta_s[2(x-x_a)/(x_b-x_a)-1], \label{eq:2}
\end{equation}
where $\tilde\Theta_s(x)$ is a dimensionless smooth step-function: 
\begin{equation}
  \label{eq:Ftr}
\tilde\Theta_s(x) = 
\begin{cases}
0, \qquad \text{for}\ x < -1, \\
\bigl[70 \sin{\left ({\pi x}/{2} \right )} + 5 \sin{\left ({3 \pi}x/{2} \right )} - \sin{\left ({5 \pi}x/{2} \right )} + 64\bigr]/128,\\
\phantom{1, \qquad {}} \text{for}\ -1 \le x < 1,\\
1, \qquad \text{for}\ x \ge 1.
\end{cases}
\end{equation}
We also introduce its integral, $\tilde\Psi_s(x) \equiv \int_{-1}^x\tilde\Theta_s(x'){\rm d}x'$,
\begin{equation}
  \label{eq:Fangle}
\tilde \Psi_s(x) = \begin{cases}
0, \qquad x < -1\\
\left[- 35 \cos{\left ({\pi x}/{2} \right )} -{5} \cos{\left ({3 \pi x}/{2} \right )}/{6} + \cos{\left ({5 \pi x}/{2} \right )}/{10}\right]/{32 \pi}\\ 
\qquad \qquad + (x+1)/2, \qquad -1 \le x < 1,\\
x, \qquad x \ge 1.
\end{cases}
\end{equation}
We use eqs~\eqref{eq:Ftr} and \eqref{eq:Fangle} to define smooth versions of the $\min$ and $\max$ functions,
% and the corresponding dimensional function $\Psi_s(x,x_0,{\rm d}x,y_0) = y_0 - {\rm d}x \tilde\Psi_s[-(x-x_0)/{\rm d}x]$. 
\begin{align}
  \label{eq:3}
  \mathrm{min}_s(f_1,f_2,{\rm d}f) &= f_2 - \tilde \Psi_s\left(\frac{f_2-f_1}{{\rm d}f}\right) {\rm d}f\\
  \mathrm{max}_s(f_1,f_2,{\rm d}f) &= -\mathrm{min}_s(-f_1,-f_2,{\rm d}f).
\end{align}
These functions are useful for introducing smooth radial breaks. 

We now use the machinery developed above to construct the mapping from the internal coordinates, $x^{(1)}, x^{(2)}\in[-1,1], x^{(3)}\in[0,2\pi]$, to the physical coordinates, $r\in[r_{\rm in},r_{\rm out}], \theta\in[0,\pi], \varphi\in[0,2\pi]$. % Since the grid is equatorially symmetric, for simplicity we focus on the northern hemisphere, $0 \le \theta < \pi/2$, or $-1\le x^{(2)} < 0$.
Quantitatively, we describe the grid as follows, and we give a qualitative explanation below:
\begin{align}
  \label{eq:grid}
  r_{\rm 1,\#} &= {\rm min}_s\left[r,r_{\rm decoll,\#},0.5r_{\rm decoll,\#}\right]/r_{\rm uniform},\\
  r_{\rm 2,\#} &= {\rm min}_s\left[r/(r_{\rm 1,\#} r_{\rm uniform}),g_{\rm \#},0.5g_{\rm \#}\right],\\
  \theta_{\rm \#} &= \pi/2+\tan^{-1}\left[r_{\rm 1,\#}^{\alpha_{1,\#}}r_{\rm 2,\#}^{\alpha_{2,\#}}\tan(x^{(2)}\pi/2)\right],
\end{align}
where ``$\#$'' stands for either ``disk'' or ``jet'' and $g_{\rm \#} = r_{\rm coll,\#}/r_{\rm decoll,\#}$. We now define a ``jet weight'', $w_{\rm jet} = \Theta_s\left(|x^{(2)}|,f_{\rm disk},f_{\rm jet},0,1\right)$, that controls the contribution of the jet grid patch for a given value of $x^{(2)}$. The values of $f_{\rm disk}$ and $f_{\rm jet}$ control the fractions of the grid devoted to resolving the disk and jet regions. We obtain the net polar angle as a function of $x^{(i)}$ via
\begin{equation}
  \label{eq:4}
  \theta = w_{\rm jet}\theta_{\rm jet} + (1-w_{\rm jet}) \theta_{\rm disk}.
\end{equation}
We direct about $25$\% of the resolution into the disk regions, $f_{\rm disk} = 0.25$, and about $40$\% of resolution into the jet regions, $f_{\rm jet} = 0.4$; the rest $35$\% resolves the disk outflow sandwiched between the disk and the jet. At $r = r_{\rm uniform}$, the angular grid is uniform. We take this radius to be equal to the inner radius of the grid, $R_{\rm in}$: uniform angular grid at the inner boundary maximizes the simulation time step and reduces time to solution. 

The value of $\alpha$ controls the collimation of the grid: in the polar regions, $\sin\theta\propto r^{-\alpha}$, and in the equatorial regions, $\cos\theta\propto r^{\alpha}$. Therefore, $\alpha>0$ leads to collimation (toward the poles) of the radial grid lines. Very close to the black hole, we choose a negative $\alpha$ value for both disk and jet regions, $\alpha_{\rm 1,disk} = \alpha_{\rm 1,jet} = -1$, which leads to a mildly decollimating grid that follows $r\cos\theta = z = {\rm constant}$ in the equatorial region and focuses the resolution on the turbulent accretion disk. (Another possible approach to focus the resolution on the equatorial plane could be to reduce the value of $r_{\rm uniform}$, but this would make the grid non-uniform at $R_{\rm in}$, reduce the time step, and increase the simulation cost.) At larger radii, at $r \gtrsim r_{\rm decoll,disk} = r_{\rm decoll,jet} = 2R_{\rm in}$, we choose $\alpha = \alpha_{\rm 2,disk} = \alpha_{\rm 2,jet} = 3/8$. This leads to a collimating grid near the poles, as seen in Fig.~\ref{fig:ic_grid}. At $r\gtrsim r_{\rm coll,disk} = 5R_{\rm in}$ the grid becomes radial in the disk region, as seen in Fig.~\ref{fig:ic_grid}. The grid becomes radial in the polar regions at much larger radii, $r \gtrsim r_{\rm coll,jet} =  10^3r_g$, well outside of Fig.~\ref{fig:ic_grid}.

{\it Cylindrification.} In standard 3D MKS coordinates, hyper-exponential MKS coordinates, and in general for all spherical-type 3D coordinate systems, the small physical extent of the cells closest to the poles introduces a severe constraint on the time-step via the Courant-Friedrichs-Lewy condition (roughly $\Delta t \lesssim \min[\Delta x]/c$, where the minimum is taken over the entire grid and in each spatial direction).  To avoid this, we use the technique of ``cylindrification'' \citep{Sasha2011}, which takes the polar cell closest to the poles and expands it laterally below a certain radius. This effectively makes the polar regions more closely resemble cylindrical coordinates than spherical. This causes the row of cells closest to the pole to be wider than the rest at $r\lesssim10r_g$, as seen in Figure~\ref{fig:ic_grid}. Because this grid deformation is concentrated at small $r$ and $\theta$ values, where the jets carry very little energy flux (enclosed jet energy flux scales as $\sin^4\theta$) and the radial velocity is directed into the black hole, this does not noticeably affect the simulations. However, it substantially -- by an order of magnitude -- speeds up the simulations.

\subsection{Density Floors}
Black hole magnetospheres naturally develop highly magnetized polar regions: gas drains off the magnetic field lines into the black hole due to gravity or is flung out to infinity due to magnetic forces. Eventually, vacuum regions would develop, which would pose numerical difficulties with grid-based MHD codes. Because of this, all such codes employ numerical floors that prevent density and internal energy from becoming too low.
When the fluid density or total internal energy dip below the floor limits, $\rho_{\rm floor} = \max[b^2/50,10^{-6}(r/r_g)^{-2}]$ or $u_{\rm floor} = \max[b^2/2500,10^{-8}(r/r_g)^{-2\gamma}]$, we add mass or energy in the drift frame, respectively. This contrasts with the more standard approaches of adding gas in the fluid frame \citep{Gammie2003} or in the zero angular momentum observer (ZAMO) frame \citep{2012MNRAS.423.3083M}. By preserving the component of the fluid momentum along the magnetic field, our drift frame floor approach (i) prevents the parallel (to the magnetic field) velocity from running away, which happens in the fluid frame floor approach (this problem becomes especially severe in 3D), (ii) avoids artificial drag on the magnetic field lines, which happens in the ZAMO frame floor approach.

Although both the ZAMO and drift frame floors result in stable numerical evolution, there are several practical advantages of the drift floor. The ZAMO floor approach is not analytic and requires an additional inversion of conserved to primitive quantities per cell per Runge-Kutta sub-step. We found that this adversely affects the speed and parallel scaling of the code: most of the floor activations occur near the black hole and thus disproportionately affect only a few MPI processes; this slows down the entire code, especially at late times in the evolution when the region affected by the floors grows in size. Second, the drift frame floors add just the right amount of mass, energy and momentum to get $\rho = \rho_{\rm floor}$ and $u_g = u_{\rm floor}$. This is not guaranteed by the ZAMO floors because of their iterative nature.

We define the normal observer frame by $\eta_\mu = (-\alpha, 0,0,0)$ where $\alpha = (-g^{tt})^{-1/2}$ is the lapse. The conserved fluid momentum in the normal observer frame is $Q_\mu \equiv -(T_g)^\nu_{\mu}\eta_\nu/\alpha  = w_g u^t u_\mu + \delta^t_\mu P_g$, where $w_g = \rho + u_g + P_g$ is the gas enthalpy \citep[see also][]{2006ApJ...641..626N}. We define the normal observer frame magnetic field 4-vector, $B^\mu = -{}^*\!F^{\mu\nu}\eta_\nu/\alpha$ and its covariant component $B_\mu = g_{\mu\nu}B^\nu$ and magnitude $B = (B^\mu B_\mu)^{1/2}$. Note that by definition, $B^t = 0$. We demand that the projection of the momentum along the magnetic field remains constant as we apply the floors:
\begin{equation}
  \label{eq:ppar}
  {\rm constant} = B^\mu Q_\mu \equiv B^i Q_i = (B^\mu v_\mu) w_g (u^t)^2 = B v_{||} w_g (u^t)^2,
\end{equation}
where $i$ runs through the spatial components only ($i=1,2,3$), $v^\mu \equiv u^\mu/u^t$ is the 3-velocity four-vector, and $v_{||} = B^\mu v_\mu /B$  is the parallel velocity component.
From eq.~\eqref{eq:ppar} we see that because the floors increase the enthalpy, $w_g$, in order to preserve the parallel momentum, we need to decrease the parallel velocity, $v_{||}$.
For convenience, we decompose $v^\mu$ into the sum of the drift and parallel velocity components, 
\begin{equation}
\label{eq:vdr}
v^\mu = v_{\rm dr}^\mu + v_{||} B^\mu/B,
\end{equation}
where the drift velocity is perpendicular to the magnetic field, 
\begin{equation}
\label{eq:Bv}
B_\mu v_{\rm dr}^\mu = 0. 
\end{equation}
Equations~\eqref{eq:vdr} and \eqref{eq:Bv} define the drift velocity and the parallel velocity.
By requiring that the gas velocity is physical, $u^\mu u_\mu = -1$, and making use of eq.~\eqref{eq:Bv}, we can express $u^t$ in terms of the parallel velocity:
\begin{equation}
  \label{eq:utsq}
  (u^t)^2 = \frac{1}{(u^t_{\rm dr})^{-2} - v_{||}^2}.
\end{equation}
Plugging this into eq.~\eqref{eq:ppar} and solving a quadratic equation for $v_{||}$ gives:
\begin{equation}
  \label{eq:vpar}
  v_{||} = \frac{x}{1+(1-x^2)^{1/2}}\times\frac{1}{u^t_{\rm dr}},
\end{equation}
where $x = 2B^\mu Q_\mu/(B w_g u^t_{\rm dr})$. After applying the floors on $\rho$ and $u_g$, we update $w_g$ and recompute $x$. Then, using eq.~\eqref{eq:vpar}, we recompute the value of $v_{||}$ and, using eq.~\eqref{eq:vdr}, the fluid velocity. As discussed above, this reduces $v_{||}$, and leads to enhanced stability of the method.
%In order to suppress unphysical electron heating in the floored regions, we introduce a passive scalar that is set to 1 whenever a floor on density or total gas internal energy is applied.  The passive scalar is evolved by being advected around with the fluid flow.  If this passive scalar is greater than $0.6$, we set $f_e =0$ and do not heat the electrons.  Furthermore, we also smoothly set $\alpha_e =0$ in these regions for conduction runs. 

\begin{figure}
    \includegraphics[width=\columnwidth]{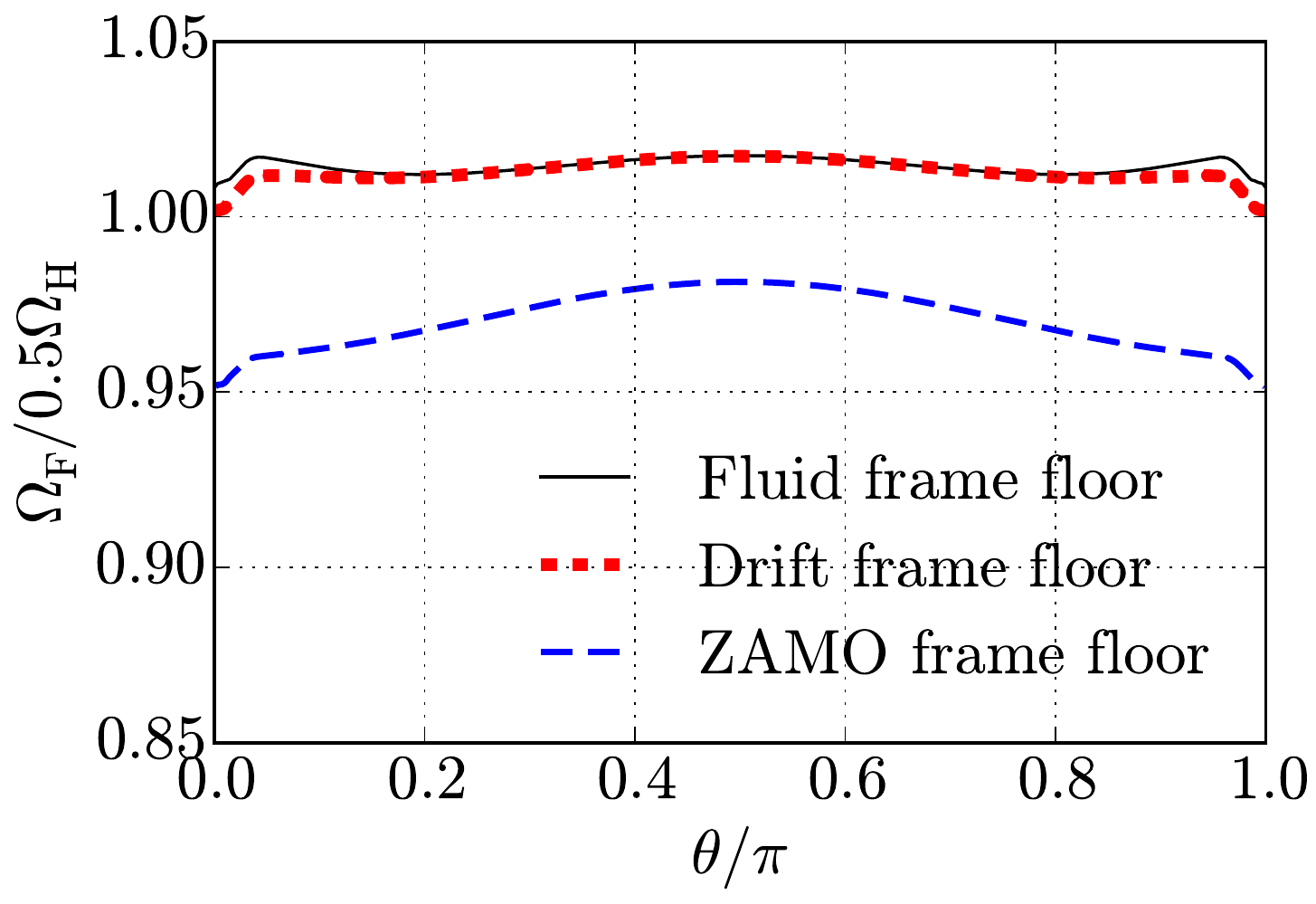} 
\caption{Angular rotational frequency $\Omega_{\rm F}$ of magnetic field lines in units of the force-free expectation, $\Omega = 0.5\Omega_{\rm H} = ac/4r_{\rm H}$, versus the polar angle for the monopolar magnetosphere test. The measurements are performed at the black hole event horizon. We expect that gas inertia makes a difference on the order of $\rho c^2/b^2 = 2$\%. The simulations with the fluid frame floor (thin solid black line) and drift frame floor (thick red short-dashed line) are within $2$\% of the analytic expectation, in line with what we expect for the small but finite gas inertia effects. For the simulation with the ZAMO frame floor (blue long-dashed line), the deviation from the analytic expectation is a bit higher, about $5\%$, with the difference especially pronounced near the poles. A larger deviation might be expected because, unlike fluid and drift frame floors, the ZAMO floors add transverse (to the magnetic field lines) momentum and therefore apply a drag to magnetic field lines.}%
\label{fig:omegaf}
\end{figure}

\emph{Floor test.} In order to verify the performance of the floor, we have simulated a monopole magnetosphere of a spinning black hole. We chose the same spin $a = 0.5$ as in our fiducial simulation and adopted a resolution of $768\times512\times1$. We used a uniform angular grid and a logarithmic radial grid with $R_{\rm in} = 0.7r_{\rm H}$ and $R_{\rm out} = 10^3r_g$. We set density and internal energy floors as $\rho_{\rm floor} = b^2/50$ and $u_{\rm floor} = b^2/2500$. Based on simulations of force-free (infinitely magnetized, $\rho, u_g\to 0$ limit) magnetospheres we expect our solution to approach $\Omega_{\rm F} = 0.5\Omega_{\rm H}$ if the floors were absent \citep{2010ApJ...711...50T}. For finite floors, however, we expect the gas inertia effects to have an effect of order $\rho_{\rm floor}/b^2 = 0.02$.

To verify this, we ran our simulation until $t = 1.7\times10^3r_g/c$, which is sufficiently long for the near black hole solution to reach a steady state and yet be unaffected by potential reflection from the outer radial grid boundary. Figure~\ref{fig:omegaf} shows the ratio of the angular rotational frequency of the magnetic field to that expected in an infinitely magnetized magnetosphere \citep{2010ApJ...711...50T}, evaluated at the event horizon. The results of the drift frame floor simulation, shown with the short dashed red line, and the fluid frame floor simulation, shown with the thin black solid line, are in line with the expectation, within $2\%$ of the force-free result. The angular frequency in the ZAMO frame floor simulation shows a larger deviation, $\simeq 5$\%, relative to the force-free value. A larger deviation might be expected because, unlike fluid and drift frame floors, the ZAMO floors add transverse (to the magnetic field lines) momentum and therefore apply a drag to magnetic field lines. This effect might contribute to the lower values of $\Omega_{\rm F}<0.5\Omega_{\rm H}$ in the polar regions reported by \citet{2012MNRAS.423.3083M}. 

\section{Motivation For the Choice of Magnetization Cut-off}
\begin{figure}
\includegraphics[scale = .105]{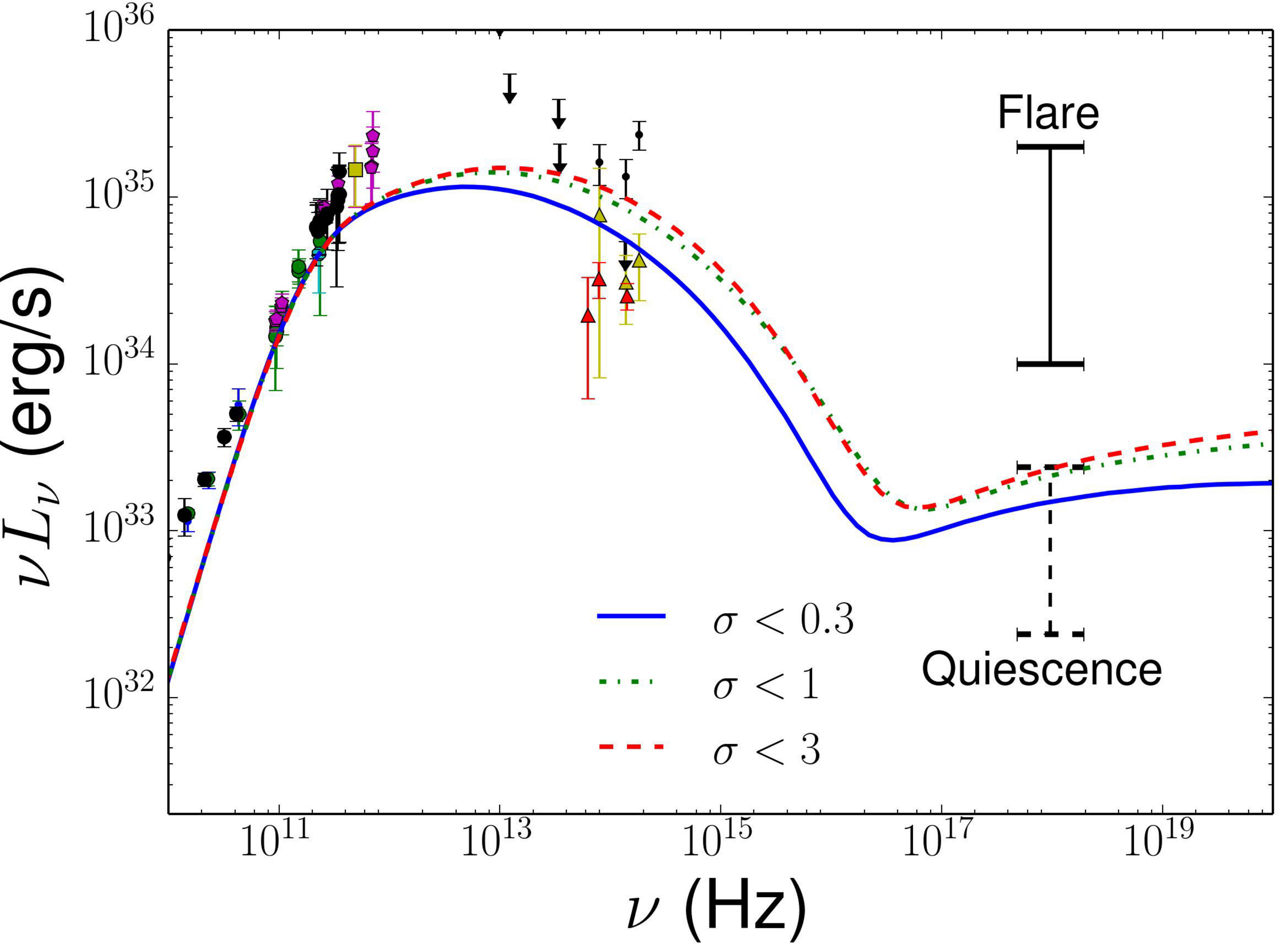}
\caption{Time-averaged SED for different limits on the magnetization parameter, $\sigma \equiv b^2/(\rho c^2)$, where regions with $\sigma < \sigma_{\rm max}$ have been excluded from the spectral calculation.  The choice of $\sigma_{\rm max}$ has no effect on the radio and millimetre emission but directly affects the NIR and X-ray emission. More restrictive cuts have reduced high frequency emission, while less restrictive cuts have increased high frequency emission.  The change in the predicted luminosity is only a factor of few for this range of $\sigma_{\rm max}$, which demonstrates that our results do not strongly depend on the choice of $\sigma_{\rm max}$.  All of the results presented in the main text use $\sigma_{\rm max} = 1$, which excludes the regions with the most uncertain thermodynamics.  }
\label{fig:sig_cut}
\end{figure}

In all of the radiative calculations shown in this work, we have limited the emitting domain of the simulation to include only the regions of the flow with $\sigma = b^2/(\rho c^2) < 1$.  The motivation for this restriction is to exclude regions which are known to have larger errors in the thermodynamics in conservative codes.  This is because as $\sigma$ approaches $\sim 1$ and above, the total energy of the fluid (which is conserved to machine precision) becomes dominated by magnetic energy, so that small errors in the magnetic field evolution lead to large errors in the internal energy of the gas.  Thus, we limit the radiative domain to regions with $\sigma < \sigma_{\rm max}$ with $\sigma_{\rm max} =1$, which is a somewhat arbitrary but conservative value chosen to minimize the effect of errors at high $\sigma$ on our results.  Figure \ref{fig:sig_cut} demonstrates how the particular choice of $\sigma_{\rm max}$ affects our results.  Over a factor of 10 in $\sigma_{\rm max}$, there is only a factor of few change in the higher frequency flux, which is encouraging for the qualitative results in this paper.   Ultimately more accurate methods will be required to model the polar region thermodynamics (where $\sigma \gtrsim 1$) more accurately.

\label{App:sigma}

\section{Steep Entropy Wave Test}

\label{sec:entwave}

In the main text we argued that numerical diffusion of entropy at steep entropy gradients is the cause of the unphysically negative heating rate near the funnel wall shown in Figure \ref{fig:2Dweight}. Here we present a simple 1D test that shows that contact discontinuities and unresolved gradients do indeed produce negative heating, as computed with our prescription for $Q$ using the local Lax-Friedrichs Riemann (LLF) solver in {\tt HARM}.  We stress that this is simply another manifestation of the unavoidable diffusion of contact discontinuities in finite-volume codes. This diffusion may be amplified by the LLF Riemann solver, which is known to be significantly more diffusive than other Riemann solvers (see, e.g., Chapters 5 and 6 in \citealt{Toro2009}).

We initialize the unmagnetized fluid with a constant velocity, $v$, and pressure, $p$, as well as either a discontinuous, square density profile, $\rho_{dis}(x)$, or a continuous profile, $\rho_{cont}(x)$, defined as:
\[ \rho_{dis}(x) = \left\{ \begin{array}{ll}
      \rho_{\rm min} & x\leq 0.4L \\
      \rho_{\rm max} & 0.4L < x < 0.6L  \\
      \rho_{\rm min} & 0.6L \leq x 
      \end{array}  \right.
\]
and
\begin{eqnarray*}
\rho_{cont}(x)  =  \left(\frac{\exp\left[ (x- 0.4L)/w\right]}{\exp\left[ (x- 0.4L)/w\right] +1} - \frac{\exp\left[ (x- 0.6 L)/w\right]}{\exp\left[ (x- 0.6L)/w\right] +1} \right) &\\ \times (\rho_{\rm max}-\rho_{\rm min}) + \rho_{\rm min}, &
\end{eqnarray*} 
where $\rho_{\rm max} = 10^{3}\rho_{\rm min}$, $w = 0.01L$, and $L$ is the size of the 1D computational domain. $\rho_{cont}$ is simply a smooth approximation to a square wave that transitions from $\rho_{\rm min}$ to $\rho_{\rm max}$ and vice-versa over a region of length $\sim w$.  The velocity is chosen to be $10^{-3} c$, which is $\approx 0.2$ of the minimum sound speed.  The pressure is chosen to be $0.012 \rho_{\rm min} c^2$, making the sound speed $\ll c$.  The electron entropy is initialized such that the electron internal energy is equal to the gas internal energy and the electron heating fraction, $f_e$, is set to either 0 or 1 to isolate the effect of the heating rate on the electron evolution.  Furthermore, the adiabatic index of the gas is $5/3$ while the adiabatic index of the electrons is $4/3$.  No floors are used on the electron variables to allow the electron entropy variable to become negative.   The simulation is performed in a periodic box and run for a single advection time, $L/v$.  The analytic solution is that the final state should be identical to the initial state and the heating rate should be 0.  We define a resolution parameter, $RP$, as the approximate number of cells per order unity change in $\kappa_g = p/\rho^\gamma$: $RP \equiv \kappa/ \Delta \kappa_g$, where $\Delta \kappa_g$ is the change in $\kappa_g$ across one cell. $RP \gg 1 $ indicates a well resolved gradient, while $RP \lesssim 1 $ represents a poorly resolved gradient. 

This set-up is chosen to mirror the typical angular entropy profiles seen in our global accretion disc simulations. These generally have a near discontinuity in entropy (and density) close to the disc-jet boundary (Figure \ref{fig:SANEFluid}).  The sharp gradient is nearly, though not perfectly, perpendicular to the grid, and the flow does have a small velocity component misaligned with the gradient.  The fact that these velocities are small (compared to the local fast magnetosonic speed used in determining the CFL condition) means that diffusion errors can be significant.  We performed a more complicated test of a contact discontinuity slightly misaligned with the grid in 2D with a thermal pressure gradient balancing a magnetic pressure gradient but found similar results to the simpler 1D hydrodynamic test; thus we focus on the latter here.  

\begin{figure}
\includegraphics[scale = .105]{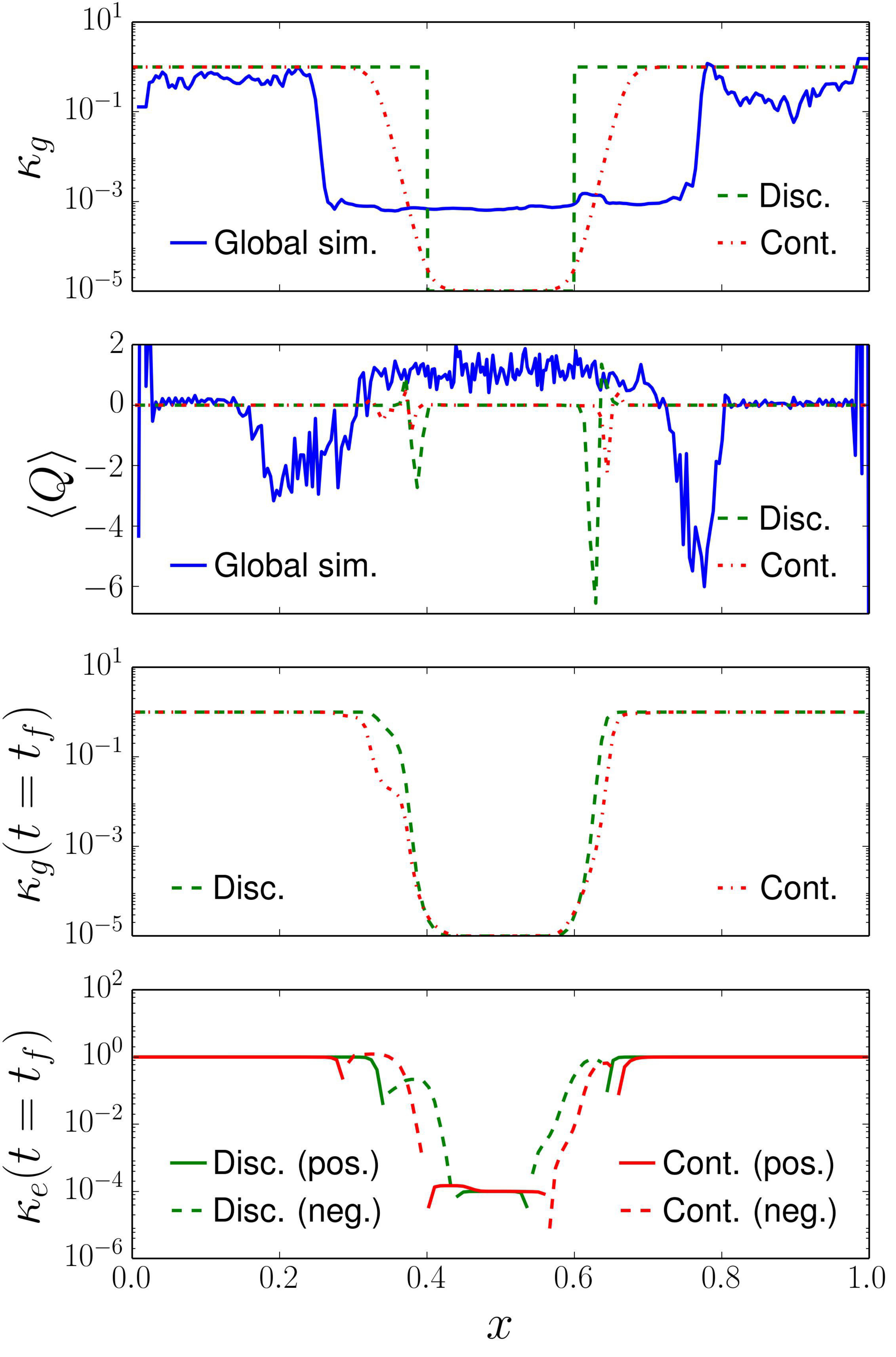}
\caption{\emph{Top Panel}: Gas entropy in our global accretion disc simulation as a function of polar angle at $r \approx 15 r_g$ at a snapshot in time and slice in $\varphi$ compared to the gas entropy initial conditions for both the discontinuous and continuous steep entropy wave test. The entropies have been scaled for ease of comparison.  \emph{Second panel}:  Time averaged Lagrangian heating rates per unit volume for both the discontinuous and continuous steep entropy wave test at a resolution of $N=128$ as well as the time and $\varphi$ averaged heating rate per unit volume in our global accretion disc simulation at $r \approx 15 r_g$ (for the global simulation, $x = \theta/\pi$). The test problem heating rates are plotted in units of $5 \times 10^{-6} \rho_{\rm max}   c^3/L$, while the global disc simulation heating rate has been arbitrarily scaled for ease of qualitative comparison. From these plots, it is clear that diffusion of large gradients in entropy leads to negative ``heating'' rates on the side of the gradient with lower entropy (corresponding to the disc side of the boundary in the accretion disc simulation).  Note that since the accretion disc simulation has been averaged over time and $\varphi$ in an Eulerian and not Lagrangian sense, motion of the disc-jet boundary increases the angular extent of the negative heating. \emph{Third and Fourth Panels}: Total gas and electron entropy at the end of the 1D entropy wave test.  All entropies have been scaled such that $\kappa_i(x=0) = 1$.  For the electrons, the magnitude of the entropy is plotted, with solid lines representing regions with positive entropy and dashed lines representing regions with negative entropy.  The electron entropy becomes significantly negative in the cells surrounding the sharp gradient while the total gas entropy is comparatively well behaved.  This shows that the effects of numerical diffusion are amplified for the electrons via the dependence on the heating rate, $Q$.  }
\label{fig:ent_profile}
\end{figure}

Figure \ref{fig:ent_profile} shows the initial entropy for this test compared with a typical angular profile of the entropy in our accretion disc simulations, along with the resulting time averaged heating rates and total gas/electron entropy after one advection time.  Note that the 1D coordinate $x$ corresponds to $\theta/\pi$ in the global disc simulation.  Large gradients in entropy clearly lead to negative heating rates (caused by excessive numerical diffusion), which appear on the low entropy side of the gradient.  In the disc simulation this is the side closest to the disc where synchroton emissivities might be significant.  Figure \ref{fig:ent_profile} (bottom panel) shows that the negative heating rates cause the electron entropy to become largely negative in regions surrounding the entropy gradient, while the total gas entropy diffuses but is reasonably well behaved. This shows that the diffusion errors more strongly affect the electrons than the total gas.  This is because small diffusive errors in the total gas entropy can lead to large errors in the calculation of $Q$ which then acts as a source term in the electron entropy evolution. It is also clear that the smooth profile has less artificial heating at this resolution; this discrepancy only increases with resolution.

\begin{figure}
\includegraphics[scale = .1]{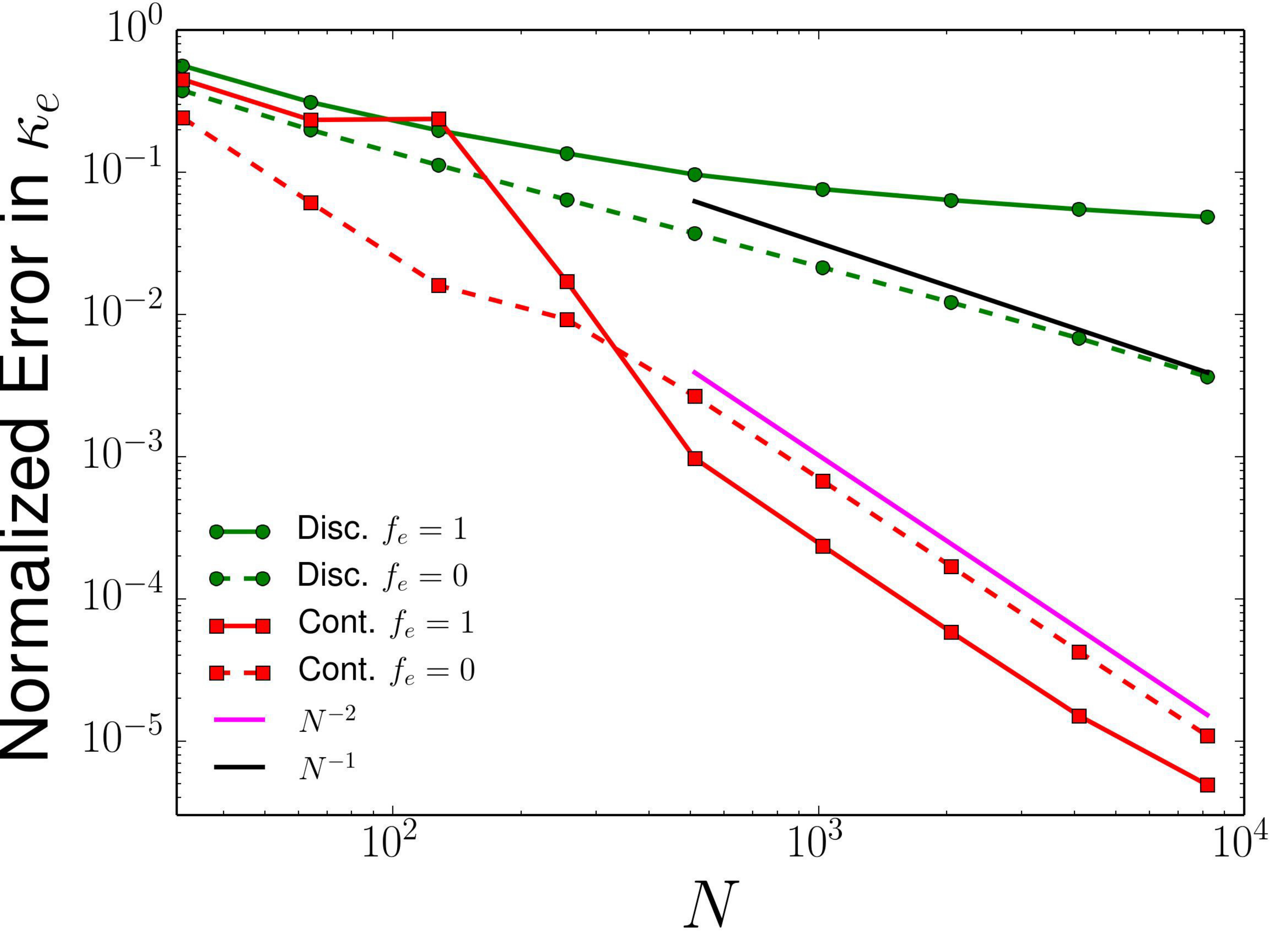}
\caption{ Error in the electron entropy after a single advection time for the steep entropy wave test, defined as $\sum |\kappa_e(t = L/v) - \kappa_e (t = 0)| / \sum \kappa_e(t=0)$.  We plot both the error in electron entropies evolved with a heating fraction of $f_e =1$ (i.e. the electrons receive all of the heat) and those evolved with a heating fraction of $f_e =0$ (i.e. the electrons are simply adiabatically advected with the flow).  As expected, when $f_e = 0$, the electron entropy converges at roughly the expected order (1st and 2nd, respectively) for the the discontinuous and continuous profiles. On the other hand, when $f_e =1$ only the continuous profile converges to the correct result.  This is because discontinuities in the flow lead to non-converging errors in the calculation of the heating rate.    Better Riemann solvers will reduce the magnitude of this error but will not improve convergence.}
\label{fig:ent_err}
\end{figure}

Figure \ref{fig:ent_err} shows the integrated error in the electron entropy at the end of the simulation normalized to the total entropy in the box as a function of resolution for the two different profiles.   The artificial heating converges to 0 at second order for the smooth profile but the artificial heating does not converge to 0 for the discontinuous profile. The reason for this is the nature of numerical diffusion.  The square wave will always be diffused over a small number of cells that is roughly independent of resolution, so the derivatives needed in the heating calculation will never be able to better resolve the gradient.  This is exactly analogous to the non-converging errors in the heating rate seen in the strong shock test of \citet{Ressler2015}.  We note that though the magnitude of the errors in Figure \ref{fig:ent_err} are relatively small at the highest resolution (with a maximum of $\sim 10 \%$ for the discontinuous wave with $f_e=1$), this depends on the chosen magnitude of the advection velocity relative to the sound speed and the duration of the 1D simulation.  The errors are larger if advection velocities perpendicular to the gradient are $\ll$ the fast magnetosonic speed.

For the smooth entropy wave test used here, the minimum resolution parameter is $RP \approx 0.42 (N/128)$, which represents the number of grid cells per order unity change in $\kappa_g$ at the steepest part of the gradient.  From Figure \ref{fig:ent_err}, we see that this means that when $RP \approx 1$, the artificial heating rate will start converging and quickly become insignificant for a smooth flow.  Two questions then arise pertaining to the accretion disc simulation:  1) How well is the gradient resolved at the fiducial resolution? and 2) Is the gradient a true discontinuity or a smooth profile that is under-resolved?  If it is in fact a true discontinuity then we have no hope of better resolving it and the negative heating will be present at all resolutions with the LLF Riemann solver.  On the other hand, if the gradient is, in fact, smooth, then higher resolution simulations will reduce the negative heating.

To attempt to answer these questions, we can directly compute the minimum resolution parameter for the entropy profile in the accretion disc simulation.   At a resolution of $320 \times 256 \times 64$ we find that the minimum resolution parameter is a mere $\sim$0.15, though at the same radius in the $640 \times 512 \times 128$ simulation the  minimum resolution parameter approximately doubles to $\sim$ 0.32.  This suggests that the gradient is smooth and that we are better resolving it with finer grids.  On the other hand, we are still under-resolving the gradient even at the very expensive resolution of $640 \times 512 \times 128$, meaning that it would take an unreasonably high resolution by today's computational standards to suppress the negative heating.   Better Riemann solvers are known to reduce the artificial numerical diffusion that we have shown leads to these negative heating rates.  Future work will explore their impact in detail.   

Finally, it is important to stress that spectrum generated by our global accretion disc simulation is reasonably well converged (Figure \ref{fig:spec_convergence}), so it is unclear whether the excess numerical diffusion and associated negative heating strongly affect the observable quantities of interest in this paper.

\label{lastpage}
\end{document}